\documentclass[aps,prc,twocolumn,superscriptaddress,showpacs,floatfix]{revtex4-2}
\usepackage[T1]{fontenc}
\usepackage[utf8]{inputenc}
\usepackage{graphicx}
\usepackage{dcolumn}
\usepackage{bm}
\usepackage{amsmath}
\usepackage{amssymb}
\usepackage{xcolor}
\usepackage{hyperref}
\usepackage{xcolor}
\usepackage{booktabs}

\begin{document}


\title{Survival of Pairing Correlations and Shell Effects at Scission in Finite-Temperature Nuclear Fission:
Implications for Odd--Even Staggering}

\author{K.~Pomorski}
\affiliation{National Centre for Nuclear Research, Pasteura 7, 02-093 Warsaw, Poland}
\author{A.~Augustyn}
\affiliation{National Centre for Nuclear Research, Pasteura 7, 02-093 Warsaw, Poland}
\author{T.~Cap}
\affiliation{National Centre for Nuclear Research, Pasteura 7, 02-093 Warsaw, Poland}
\author{Y.~J.~Chen}
\affiliation{China Institute of Atomic Energy, Nuclear Data Center, Beijing 102413, China}
\author{M.~Kowal}
\email{m.kowal@ncbj.gov.pl}
\affiliation{National Centre for Nuclear Research, Pasteura 7, 02-093 Warsaw, Poland}
\author{M.~Warda}
\affiliation{Department of Theoretical Physics, Maria Curie-Sk{\l}odowska University, 20-031 Lublin, Poland}
\author{Z.~G.~Xiao}
\affiliation{Department of Physics, Tsinghua University, Beijing 100084, China}

\begin{abstract}
We investigate the finite-temperature evolution of microscopic free-energy corrections in nuclear fission, focusing on pairing and shell effects near scission. The analysis is based on a finite-temperature BCS treatment combined with the Strutinsky method and is performed for representative deformation points along the fission path.
Both pairing and shell contributions exhibit regular thermal attenuation, but their deformation dependencies differ substantially. In particular, pairing remains strongly deformation-dependent in the scission region, and its free-energy contribution differs markedly between the constant and surface-dependent pairing-strength prescriptions. The shell correction near scission is also significant at low temperature and is progressively suppressed with increasing excitation energy.
These results support the interpretation of odd--even staggering in fragment charge yields as a manifestation of pairing correlations surviving into the strongly deformed pre-scission configuration. They also show that pairing and shell effects should be treated separately in finite-temperature dynamical calculations, with distinct deformation- and temperature-dependent attenuation laws.
\end{abstract}


\maketitle


\section{Introduction}
\label{sec:intro}

It is well established that pairing correlations affect fission through two closely related mechanisms: they modify the deformation energy landscape and, even more importantly, they strongly reduce the collective inertia associated with large-amplitude motion~\cite{Brack1972,Ledergerber1973,Moretto1974,Staszczak1985,Lojewski1980,Staszczak1989,Lojewski1999,Giuliani2014,Sadhukhan2014,Zhao2016,RodriguezGuzman2023}. 
As a consequence, enhanced pairing may favor dynamical fission paths that differ substantially from the static minimum-energy trajectory and may reduce predicted fission lifetimes by several orders of magnitude~\cite{Moretto1974,Staszczak1985,Staszczak1989,Giuliani2014,Sadhukhan2014}.

Pairing correlations influence fission not only through their contribution to the deformation energy and collective inertia, but also through their impact on the microscopic mechanism by which the system evolves across the dense sequence of avoided single-particle level crossings between the ground state and scission. In the mean-field picture, the collective motion may proceed either adiabatically, with the occupied orbitals adjusting to the instantaneous lowest-energy configuration, or diabatically, with particles tending to preserve their original single-particle character through the crossing~\cite{Landau1932,Zener1932,Stueckelberg1932,HillWheeler1953,SchutteWilets1978,Nazarewicz1993,KowalSkalski2022}. Since adiabatic evolution is favored by stronger residual interaction and slower collective motion, while weaker interaction and faster motion enhance diabatic continuation, pairing may be viewed as one of the key microscopic ingredients promoting adiabaticity along the fission path. Conversely, when pairing is reduced, the evolution becomes more susceptible to diabatic effects, which may modify the effective barrier.

The role of pairing in nuclear fission has recently been revisited in a systematic microscopic study by Zdeb \textit{et al.}~\cite{Zdeb2025}, where the impact of pairing on fission barriers, collective inertias, and spontaneous-fission lifetimes was analyzed within a self-consistent framework. That work also provides a broad and useful overview of the literature on pairing in fission, from early macroscopic-microscopic treatments to modern energy-density-functional approaches.

At the same time, recent Langevin-based calculations of isotope-resolved fission yields have shown that, although the dominant isotopic trends and heavy--light fragment correlations are reproduced rather well, the odd--even staggering observed in charge-resolved yields remains systematically weaker in the calculations than in the evaluated data~\cite{Pomorski2024LM,PomorskiAcceptedBaXe}. This discrepancy points to a missing or still oversimplified description of pairing-related structure in the final stage of the fission process.

The underlying question is both fundamental and unresolved: what becomes of pairing correlations and residual interactions at scission, particularly at higher excitation energies? As pairs are broken during the descent toward rupture, the emerging unpaired nucleons may contribute additional single-particle angular momenta and intrinsic excitation, thereby affecting the angular-momentum content and its sharing between the fragments \cite{Wilson2021}. The origin of fragment angular momenta is itself still an open problem in fission, and the thermal quenching of pairing near scission may be one of its missing ingredients.

Here, we investigate the finite-temperature evolution of pairing and shell corrections to the free energy from the ground state to scission. Our aim is to establish how these corrections are attenuated with increasing temperature and whether the persistence of pairing correlations in the scission region can be linked to the survival of odd--even staggering in fragment charge yields.

More specifically, we analyze the temperature dependence of the pairing gap, the pairing free-energy correction, and the shell free-energy correction for representative deformation points, including the ground state, the second saddle, and asymmetric and symmetric scission configurations. We show that these quantities admit compact scaling representations that may be useful for finite-temperature fission modeling. We then argue that the gradual disappearance of odd--even staggering with increasing excitation energy can be naturally interpreted as a manifestation of the thermal damping of pairing correlations in the pre-scission and scission region.

In this sense, the present paper provides a microscopic finite-temperature complement to earlier Langevin studies of charge polarization and isotope-resolved fission yields~\cite{Pomorski2024LM,PomorskiXiao2025ISOLDA}. While those works established the phenomenology of charge partitioning and identified the limitations of effectively damping microscopic corrections, the present study aims to clarify the underlying pairing physics.

The paper is organized as follows. Section~\ref{sec:shape} introduces the shape degrees of freedom and the Fourier-over-Spheroid parametrization used throughout this work. Section~\ref{sec:pairing_formalism} presents the finite-temperature pairing formalism. Sections~\ref{sec:gap} and~\ref{sec:fpair} discuss the thermal behavior of the pairing gap and of the pairing free-energy correction, respectively. Section~\ref{sec:shell_damping} defines the shell free-energy correction and analyzes its damping from compact shapes up to scission. Section~\ref{sec:method} describes the stochastic Langevin framework used to connect the microscopic free-energy corrections with fragment observables. Section~\ref{sec:cheq} applies this framework to charge polarization and to the odd--even effect in fragment charge yields. The discussion and conclusions are given in Sec.~\ref{sec:discussion_conclusions}.

\section{Shape degrees of freedom and FoS parametrization}
\label{sec:shape}

The nuclear shapes considered in this work are described within the Fourier-over-Spheroid (FoS) parametrization introduced in Refs.~\cite{Pomorski2023FoS,PomorskiNerlo2023FoSNonax}. Both the finite-temperature free-energy maps and the subsequent dynamical analysis are formulated in terms of these collective variables.

In cylindrical coordinates $(\rho,\varphi,z)$, the nuclear surface is written as
\begin{equation}
\rho_s^2(z,\varphi)=\frac{R_0^2}{c}\,f\!\left(\frac{z-z_{\rm sh}}{z_0}\right)
\frac{1-\eta^2}{1+\eta^2+2\eta\cos(2\varphi)},
\label{eq:FoS_surface_main}
\end{equation}
where $R_0$ is the radius of the spherical reference shape, $z_{\rm sh}$ denotes the shift required to keep the center of mass at the origin, and the half-length of the nucleus is $z_0=cR_0$. The axial profile function is expanded in a Fourier series,
\begin{equation}
\begin{aligned}
f(u)&=1-u^2-\sum_{n=1}^{\infty}
\Bigg[a_{2n}\cos\!\left(\frac{2n-1}{2}\pi u\right) \\
&\phantom{=}\,+\,a_{2n+1}\sin(n\pi u)\Bigg], \qquad
u=\frac{z-z_{\rm sh}}{z_0}.
\end{aligned}
\label{eq:FoS_profile_main}
\end{equation}
and the collective coordinate space is defined by the four-component vector
\begin{equation}
\vec q=(c,a_3,a_4,\eta).
\label{eq:FoS_coordinates}
\end{equation}

Here $c$ controls the elongation of the system, $a_3$ describes left--right mass asymmetry, $a_4$ governs neck development, and $\eta$ parametrizes non-axiality through elliptic deformation of the transverse cross section. In this way, the FoS parametrization provides a flexible description of compact, superdeformed, necked, and reflection-asymmetric configurations within one continuous shape manifold. A particularly important feature of the FoS parametrization is that it offers a natural geometrical description of configurations approaching scission. Within this framework, the rupture configuration corresponds geometrically to the limit in which the nuclear surface touches the symmetry axis, i.e.\ $\rho(z)\to 0$ at some point along the elongation coordinate. In practice, a finite critical neck radius is adopted, which is expressed directly in terms of the deformation parameters, most notably through the neck-related coordinate $a_4$. In the present work, the scission line is defined by adopting a critical neck radius of the order of the nucleon radius, which in the FoS parametrization corresponds approximately to $a_4 \approx 0.72$. More precisely, the associated parameter $r_{\rm neck}$ is defined for a sharp, uniform density distribution, so that the corresponding neck radius for a diffuse density profile is slightly larger; More precisely, the associated parameter $r_{\rm neck}$ is defined for a sharp, uniform density distribution, so that the corresponding neck radius for a diffuse density profile is slightly larger; further details, including the extension to non-axial shapes, the volume-conservation condition, and the center-of-mass correction, are given in Refs.~\cite{Pomorski2023FoS,PomorskiNerlo2023FoSNonax}.

\begin{figure}[t]
\centering
\includegraphics[width=\columnwidth]{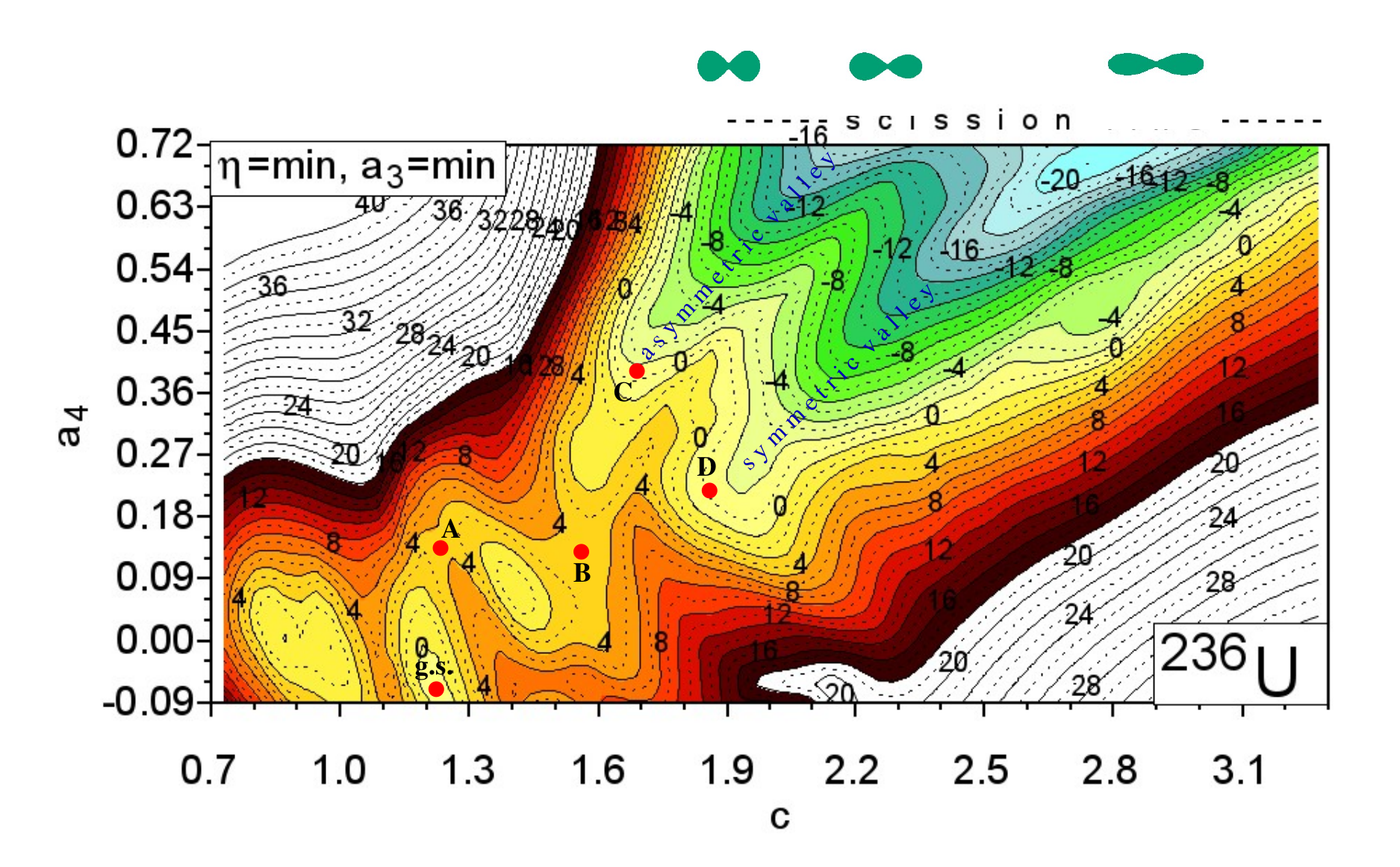}
\caption{Zero-temperature deformation-energy landscape of $^{236}$U in the $(c,a_4)$ plane, obtained by minimization over the non-axiality parameter $\eta$ and the mass-asymmetry coordinate $a_3$ at each point. Indicated on the map are the ground state (g.s.), the first saddle point $A$, the second saddle point $B$, and the superdeformed second minimum located between them. Beyond the second saddle, the collective path bifurcates toward the asymmetric valley through point $C$ and toward the symmetric valley through point $D$. The dashed line at the top denotes the adopted scission line.}
\label{fig:U236_landscape}
\end{figure}

In this work, we adopt a macroscopic--microscopic approach formulated within the FoS shape parametrization, in which the deformation energy is constructed as a sum of a smooth macroscopic contribution and microscopic shell and pairing corrections. The macroscopic part is taken from the Lublin--Strasbourg Drop (LSD) model~\cite{PomorskiDudek2003LSD}, while the microscopic shell and pairing corrections are obtained from the Yukawa-folded single-particle potential~\cite{Dobrowolski2016YukawaFolded}. The resulting zero-temperature potential-energy surface provides the reference landscape on which the subsequent discussion of pairing, shell damping, and scission configurations is based. An example of the calculated potential-energy surface is shown in Fig.~\ref{fig:U236_landscape} for $^{236}$U in the $(c,a_4)$ plane, where at each point of the map the energy is minimized with respect to $\eta$ and $a_3$. Within this representation, the characteristic configurations entering the subsequent analysis can be identified within a single set of collective coordinates.

Starting from the ground state (g.s.), the fission path first crosses the first saddle point, labeled $A$. At larger elongation, the landscape exhibits a second, superdeformed minimum, followed by the second saddle point $B$, which serves as the starting configuration for the stochastic Langevin trajectories considered later. Beyond $B$, the energy surface opens toward the scission region and separates into two competing branches. The entrance to the asymmetric valley is marked by point $C$, whereas the entrance to the symmetric valley is marked by point $D$. The upper boundary of the map is defined by the adopted scission line, which marks the onset of rupture in the FoS parametrization. In addition, several representative nuclear shapes corresponding to points along the scission line are shown above the map. They illustrate explicitly the type of near-rupture configurations generated by the FoS parametrization in this region of collective space.

\section{Finite-temperature pairing formalism}
\label{sec:pairing_formalism}

Although the fissioning nucleus is, strictly speaking, an isolated quantum system and should therefore be described within a micro-canonical framework, the pronounced separation of timescales between slow collective motion and much faster intrinsic single-particle rearrangements justifies an adiabatic treatment. In this picture, the intrinsic degrees of freedom are assumed to reach local thermal equilibrium much faster than the collective coordinates can change significantly. As a consequence, for each fixed deformation point along the fission path, the intrinsic subsystem may be described by a local canonical ensemble characterized by a deformation-dependent temperature.

Following the standard finite-temperature treatment used in many studies of pairing and shell effects in fission~\cite{Brack1972,Ivanyuk2018}, we describe pairing correlations within the Bardeen--Cooper--Schrieffer framework generalized to nonzero temperature~\cite{Bardeen1957}. At a given deformation, this formalism provides the thermal occupation of quasiparticle states and the corresponding paired energy functional~\cite{Ivanyuk1999,Ivanyuk2018}

\begin{equation}
E_{\rm BCS}(T)=2\sum_{\nu>0} e_{\nu} n_{\nu}^{\Delta,T}-\frac{\Delta(T)^2}{G}
- G\sum_{\nu>0}v_{\nu}^4,
\label{eq:EBCS}
\end{equation}
where $E_{\rm BCS}(T)$ is the finite-temperature BCS energy of the paired system, $e_{\nu}$ denotes the single-particle energy of the state $\nu$, and the summation over $\nu>0$ runs over one representative from each time-reversed pair of states. The factor 2 accounts for the Kramers degeneracy of such pairs. The quantity $G$ is the pairing-strength parameter, $\Delta(T)$ is the temperature-dependent pairing gap, and $v_\nu$ is the usual BCS occupation amplitude, with $v_\nu^2$ representing the occupation probability of the state $\nu$ in the paired solution. The thermal occupation number in the paired system is written as
\begin{equation}
n_{\nu}^{\Delta,T}=\frac{1}{2}\left(1-\frac{e_{\nu}-\lambda}{E_{\nu}}
\tanh\frac{E_{\nu}}{2T}\right),
\label{eq:occupation_pairing}
\end{equation}
where $T$ is the temperature, expressed in MeV, and $\lambda$ is the chemical potential of the paired system. The quantity
\begin{equation}
E_{\nu}=\sqrt{(e_{\nu}-\lambda)^2+\Delta(T)^2}
\label{eq:quasiparticle_energy}
\end{equation}
is the quasiparticle energy corresponding to the single-particle state $\nu$. The particle number and the pairing gap are determined self-consistently from the coupled finite-temperature BCS equations
\begin{equation}
\begin{aligned}
N &= 2\sum_{\nu>0} n_{\nu}^{\Delta,T}, \qquad
\frac{2}{G} = \sum_{\nu>0}\frac{1}{E_{\nu}}
\tanh\frac{E_{\nu}}{2T},
\end{aligned}
\label{eq:number_gap_eq}
\end{equation}
where $N$ denotes the number of particles in the subsystem under consideration, i.e., either protons or neutrons.

Within the standard BCS picture, the pairing gap disappears at a critical temperature approximately given by
\begin{equation}
T_c \approx \left(\frac{e^C}{\pi}\right)\Delta(T=0)\approx 0.57\,\Delta(T=0),
\label{eq:Tc_standard}
\end{equation}
where $C$ is Euler's constant.

The pairing energy correction is defined as the difference between the BCS energy of the paired system and the energy of the corresponding independent-particle reference system:
\begin{equation}
E_{\rm pair}(T)=E_{\rm BCS}(T)-2\sum_{\nu>0}n_{\nu}^{T}e_{\nu},
\label{eq:Epair}
\end{equation}
where the Fermi--Dirac occupation numbers for the unpaired system are
\begin{equation}
n_{\nu}^{T}=\frac{1}{1+\exp[(e_{\nu}-\mu)/T]}.
\label{eq:occupation_free}
\end{equation}
Here, $\mu$ denotes the chemical potential of the unpaired reference system. It should be emphasized that the quantities $\lambda$ and $\mu$ are, in general, different. The parameter $\lambda$ is the chemical potential associated with the paired finite-temperature BCS solution and enters the quasiparticle occupations $n_{\nu}^{\Delta,T}$, whereas $\mu$ is the chemical potential of the corresponding unpaired reference system and determines the Fermi--Dirac occupations $n_{\nu}^{T}$. Since these two systems are characterized by different occupation schemes, their chemical potentials are determined independently from the corresponding particle-number conditions and need not coincide. In this sense, the present formalism involves two distinct Fermi levels, corresponding to the paired and unpaired descriptions.

At finite excitation energy, the relevant quantity for the local thermodynamic description is not the total energy alone but the free energy. This follows directly from the local canonical treatment adopted at fixed deformation: once the intrinsic subsystem is assumed to be in thermal equilibrium at temperature $T$, the entropy contribution must be included explicitly. For this reason, the microscopic pairing contribution entering the finite-temperature deformation energy should be formulated as a correction to the free energy rather than merely to the internal energy.

For the unpaired reference system, one obtains the usual Fermi-gas expression, whereas for the paired system, the entropy is defined within the finite-temperature superfluid formalism~\cite{Decowski1968,Adeev1975}:
\begin{equation}
S(T)=-2\sum_{\nu>0}\left[n_{\nu}^{T}\ln n_{\nu}^{T}
+(1-n_{\nu}^{T})\ln(1-n_{\nu}^{T})\right],
\label{eq:S_free}
\end{equation}
\begin{equation}
S_{\rm pair}(T)=-2\sum_{\nu>0}\left[n_{\nu}^{\Delta,T}\ln n_{\nu}^{\Delta,T}
+(1-n_{\nu}^{\Delta,T})\ln(1-n_{\nu}^{\Delta,T})\right].
\label{eq:S_pair}
\end{equation}
Similar finite-temperature calculations of entropies, level-density parameters, and fission-related quantities along deformation paths have recently been performed for heavy and superheavy nuclei in Refs.~\cite{Rahmatinejad2024EPJA,Rahmatinejad2021PRC}.

The effect of the pairing correlation on the free energy is then given by the difference between the free energies of the paired and unpaired reference systems at the same deformation and temperature,
\begin{equation}
F_{\rm pair}(T)=\left[E_{\rm BCS}(T)-T S_{\rm pair}(T)\right]
-\left[E(T)-T S(T)\right],
\label{eq:Fpair_def}
\end{equation}
where $E(T)=2\sum_{\nu}e_\nu n_\nu^T$ is the energy of the unpaired reference system.


\section{Temperature dependence of the pairing gap}
\label{sec:gap}

\begin{figure*}[t]
\centering
\includegraphics[width=0.48\textwidth]{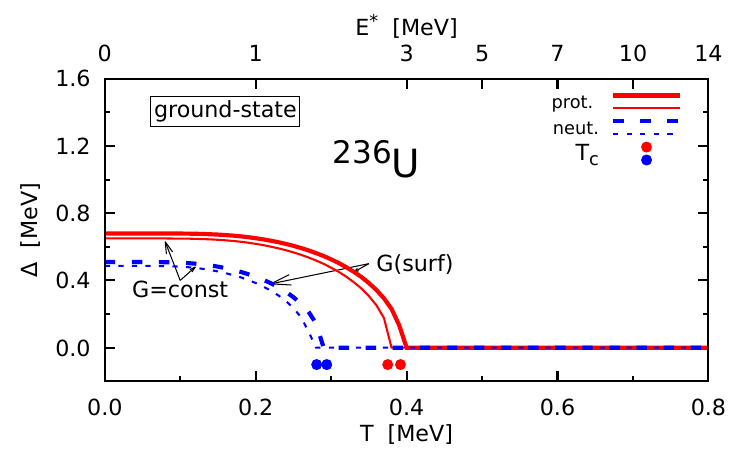}
\includegraphics[width=0.48\textwidth]{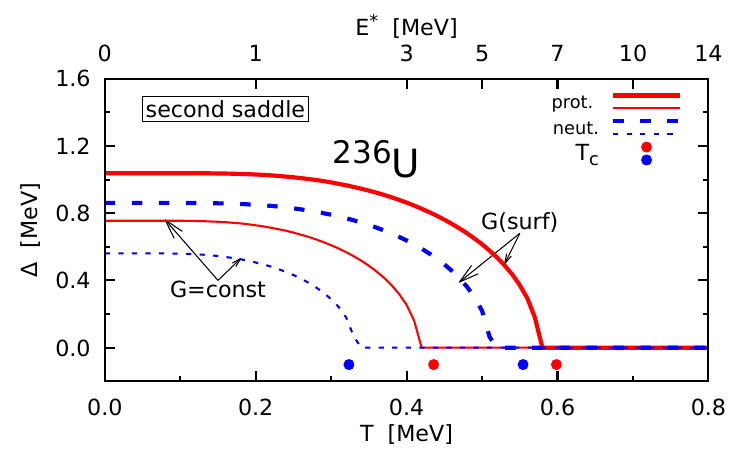}\\
\includegraphics[width=0.48\textwidth]{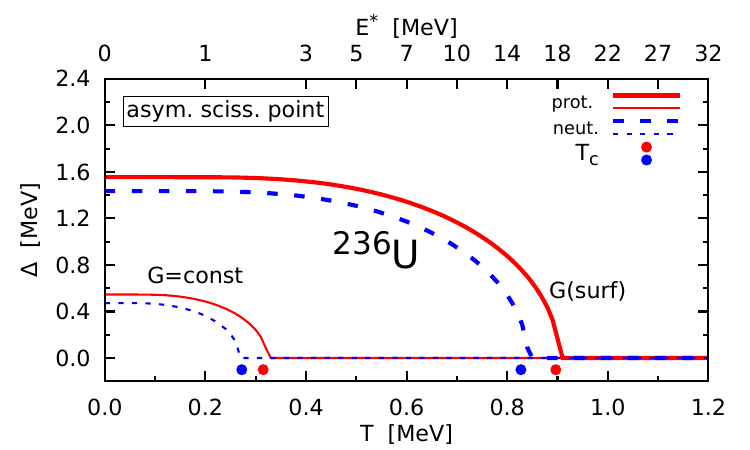}
\includegraphics[width=0.48\textwidth]{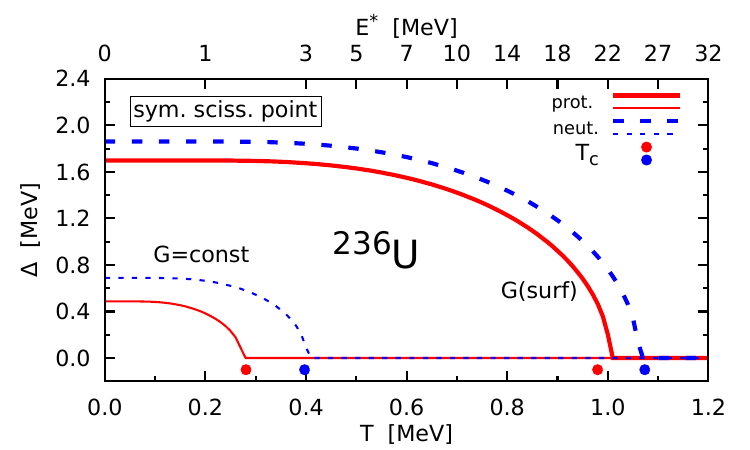}
\caption{Temperature dependence of the proton (red solid lines) and neutron (blue dashed lines) pairing gap $\Delta(T)$ for $^{236}$U at four representative deformation points along the fission path: ground state, second saddle, asymmetric scission point, and symmetric scission point. In each panel, proton and neutron gaps are shown for two pairing prescriptions, $G={\rm const}$ and $G\propto Surface$. The upper horizontal scale indicates the corresponding excitation energy. The results show that pairing correlations persist up to deformation-dependent critical temperatures and remain non-negligible even in the scission region.}
\label{fig:gap_all}
\end{figure*}

As early as the 1960s, it was argued that pairing correlations in finite nuclei are predominantly associated with the nuclear surface rather than the bulk volume~\cite{Kennedy1964,Nilsson1969}. Within that line of reasoning, if the effective pairing strength scales with the nuclear surface area, the pairing gap may exhibit a pronounced dependence on deformation due to variations in the surface-to-volume ratio. In particular, one finds in the older literature the suggestion of a rather strong scaling, $\Delta\propto (Surface)^3$, whereas in the slab-model analysis of Kennedy \textit{et al.} a weaker dependence, $\Delta\propto (Surface)^{3/2}$, was obtained~\cite{Kennedy1964}. Related surface-pairing arguments were also discussed by St\k{e}pie\'n, Szyma\'nski and by Nilsson \textit{et al.}~\cite{StepienSzymanski1968,Nilsson1969}. A related argument may also be formulated directly for the effective pairing strength. In our calculations, the pairing interaction is based on the prescription $G{\cal N}^{2/3}=g_0\hbar\omega_0$, with $g_0=0.29$ for both protons and neutrons and with ${\cal N}=Z$ or $N$, respectively. This relation was derived in Ref.~\cite{BoningSobiczewskiPomorski1985} using harmonic-oscillator arguments analogous to those employed by Bohr and Mottelson in the derivation of the shell-damping coefficient. Then the effective pairing strength should increase with deformation. This observation is again consistent with the present conclusion that the survival of pairing at large elongation is physically well motivated. Recent self-consistent calculations also support the conclusion that pairing remains an important ingredient of fission dynamics at large deformation and finite excitation energy, through its influence on collective inertia, barrier penetration, and the effective fission path~\cite{Zhao2019PRC,Zhao2021PRC,Li2015PRC,Wang2023PRC}. 

The temperature dependence of the pairing gap $\Delta(T)$, obtained from the finite-temperature BCS equations, is displayed in Fig.~\ref{fig:gap_all} for four representative configurations visible in Fig.~\ref{fig:U236_landscape}: the ground state, the second saddle, and the asymmetric and symmetric scission points. The results reveal a clear and systematic evolution with deformation, nucleon type, and the adopted pairing prescription.

In the panel corresponding to the ground-state configuration, the proton and neutron solutions obtained with $G={\rm const}$ and with $G\propto Surface$ remain relatively close to one another over the whole temperature interval. This is the configuration for which the difference between the two pairing prescriptions is smallest, and the corresponding critical temperatures are also the least separated. The thermal suppression of pairing is therefore comparatively insensitive to the precise choice of the effective interaction in the vicinity of the compact equilibrium shape. At the same time, the proton and neutron branches display their highest degree of mutual similarity in this panel, which indicates that near the ground state, the underlying single-particle structure affects both subsystems in a broadly comparable manner.

A qualitatively different situation emerges already at the second saddle. Here, the solutions begin to separate visibly, and the critical temperature associated with the surface-dependent pairing strength shifts upward relative to that obtained for the constant monopole force. The proton and neutron branches no longer remain as close to one another as in the ground-state case, indicating that, with increasing deformation, the thermal response of pairing becomes more sensitive to both the details of the single-particle spectrum and the form of the pairing interaction. The second saddle thus marks the onset of a regime in which the pairing prescription ceases to be merely a quantitative detail and begins to substantially influence the thermal persistence of superfluid correlations.

In the asymmetric scission configuration, the proton and neutron solutions again move somewhat closer to one another than at the second saddle, but the difference between the two pairing prescriptions becomes markedly larger. In particular, the solutions obtained with $G \propto Surface$ remain finite over a much broader temperature interval than those obtained with $G={\rm const}$. This shows that once the system enters the strongly elongated pre-scission region, the choice of pairing prescription has a decisive impact not only on the zero-temperature gap but also on the temperature range over which pairing correlations persist. In other words, at large elongations and developed necking, the surface-dependent interaction no longer acts as a small refinement of the constant-$G$ scheme; it leads instead to a qualitatively different picture of the thermal evolution of the superfluid field.

This effect is most dramatic in the symmetric scission configuration, where the separation between the two pairing prescriptions is largest. In this case, the critical temperatures are strongly displaced with respect to one another, and the two models imply entirely different scenarios for the thermal quenching of pairing correlations near the symmetric scission line. The surface-dependent pairing prescription predicts that pairing survives up to substantially higher temperatures, whereas the constant monopole interaction leads to a much earlier collapse of the gap. The difference is therefore not merely quantitative. It changes the physical interpretation of the scission region itself, because it determines whether pairing should still be regarded as an active microscopic ingredient at the stage where the final fragment properties are being established.

The strong deformation dependence visible in Fig.~\ref{fig:gap_all} may also be viewed in the broader context of the surface character of nuclear pairing. The difference between the constant-$G$ and surface-dependent prescriptions is small near the ground state, where the nucleus remains comparatively compact, and the effective surface modification is modest. By contrast, the difference grows rapidly with elongation and becomes particularly pronounced in the scission region, where the geometry is dominated by a strongly developed neck and by a much larger effective surface. This trend is exactly what one would expect if pairing were increasingly controlled by surface properties as the system evolves toward rupture.
\begin{figure}[b]
\centering
\includegraphics[width=\columnwidth]{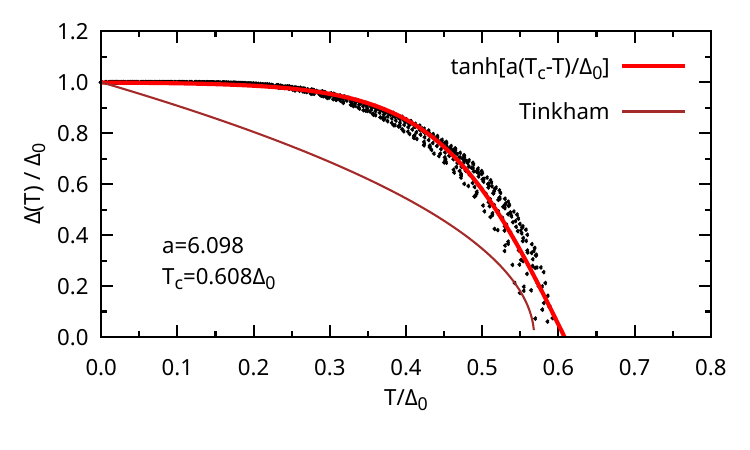}
\caption{Scaling representation of the temperature dependence of the pairing gap. The black points collect all calculated $\Delta(T)/\Delta_0$ values shown in Fig.~\ref{fig:gap_all}, including proton and neutron systems, two pairing-strength prescriptions, and four representative deformation points. The red curve corresponds to the fit $\Delta(T)=\Delta_0\tanh[a(T_c-T)/\Delta_0]$ with $T_c=0.608\Delta_0$ and $a=6.098$, while the thin reference curve shows the standard approximation valid only near the critical temperature.}
\label{fig:Delta_scaling}
\end{figure}
The global shapes of all curves remain strikingly similar despite their large differences in absolute scale. In every panel, one observes the same general pattern: a weak temperature dependence at low $T$, followed by an accelerated decrease at intermediate temperatures, and finally a rapid collapse as the critical temperature is approached. This similarity is not accidental. It indicates that the thermal suppression of pairing is primarily governed by two characteristic scales: the zero-temperature gap and the corresponding critical temperature, whereas the dependence on deformation, nucleon type, and pairing prescription is mainly reflected in variations in these scales. This point is demonstrated more directly in Fig.~\ref{fig:Delta_scaling}, where the individual solutions, once expressed in reduced variables, collapse onto a common average dependence. In this sense, the four panels of Fig.~\ref{fig:gap_all} do not merely document the sensitivity of the gap to shape and interaction; together with Fig.~\ref{fig:Delta_scaling}, they point to a near-universal scaling behavior of pairing quenching once the problem is expressed in properly reduced variables.

This conclusion is physically important. It means that even though the absolute magnitude of the pairing gap and the location of the critical temperature can differ dramatically between the ground state, the saddle region, and the scission configurations, the underlying mechanism of thermal damping remains essentially the same. The various deformation points and pairing prescriptions therefore generate different realizations of one and the same generic behavior, rather than qualitatively unrelated patterns. This near-universality provides strong support for using compact phenomenological scaling laws to represent the finite-temperature suppression of pairing correlations along the fission path. At the same time, the systematic increase of the scale differences toward scission demonstrates that the survival of pairing in the fragment-formation region depends crucially on the adopted pairing prescription, which may have direct consequences for the microscopic interpretation of odd--even staggering and related scission observables.

Generally, one can see that the calculated pairing gaps exhibit the expected decrease with increasing temperature and vanish near a critical temperature close to the textbook estimate of Eq.~(\ref{eq:Tc_standard}). At the same time, a comparison of all calculated cases suggests that their temperature evolution can be described by a common scaling variable. An effective representation of the average behavior is given by
\begin{equation}
\Delta(T)=\Delta_0\tanh\left(a\frac{T_c-T}{\Delta_0}\right),
\label{eq:Delta_scaling}
\end{equation}
where $\Delta_0\equiv \Delta(T=0)$ and the fitted parameters are
\begin{equation}
T_c = 0.608\,\Delta_0, \qquad a=6.098.
\label{eq:Delta_scaling_params}
\end{equation}
Equation~(\ref{eq:Delta_scaling}) is intended as an empirical scaling law suitable for practical applications. It should not be confused with asymptotic expressions derived near $T_c$ in condensed-matter superconductivity, which are only valid in the immediate vicinity of the critical point \cite{Tinkham}.


\section{Free-energy pairing correction}
\label{sec:fpair}

\begin{figure*}[t]
\centering
\includegraphics[width=0.48\textwidth]{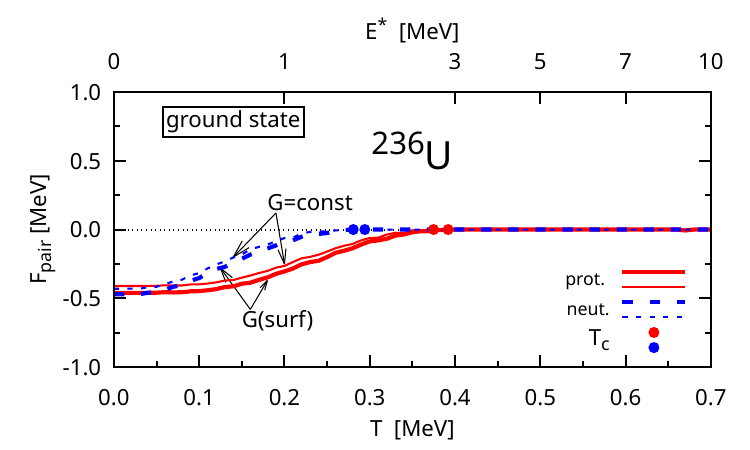}
\includegraphics[width=0.48\textwidth]{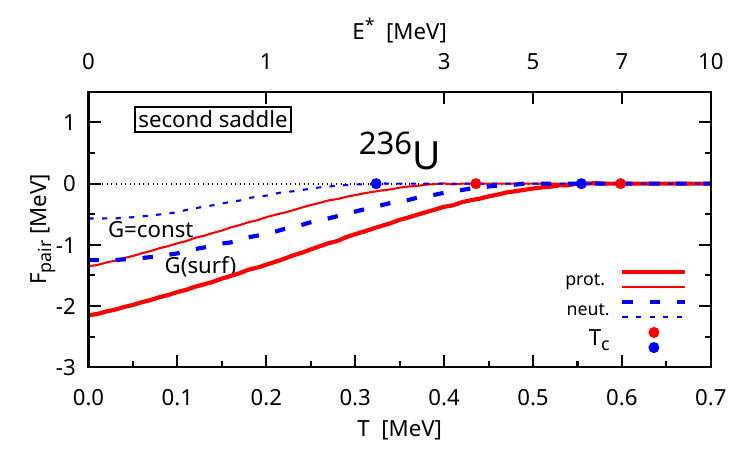}\\
\includegraphics[width=0.48\textwidth]{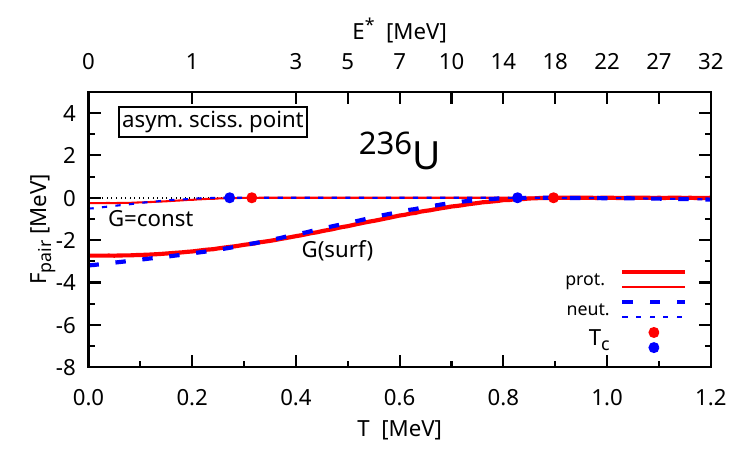}
\includegraphics[width=0.48\textwidth]{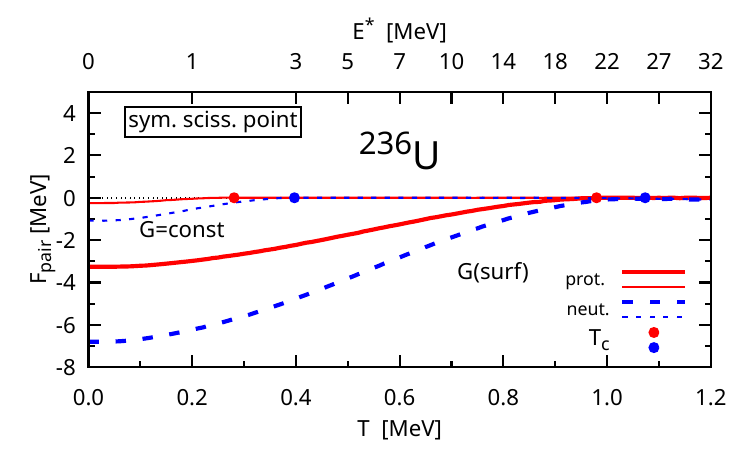}
\caption{Temperature dependence of the proton (red solid lines) and neutron (blue dashed lines) pairing free-energy corrections $F_{\rm pair}(T)$ for $^{236}$U at four representative deformation points: ground state, second saddle, asymmetric scission, and symmetric scission. As in Fig.~\ref{fig:gap_all}, both proton and neutron subsystems are shown for the two pairing prescriptions $G={\rm const}$ and $G\propto surface$. The overall trend demonstrates a smooth damping of the pairing contribution to the free energy with increasing temperature.}
\label{fig:Fpair_all}
\end{figure*}

An analogous but in some respects even more revealing picture emerges from Fig.~\ref{fig:Fpair_all}, which shows the temperature dependence of the pairing free-energy correction. In the ground-state configuration, the proton and neutron curves lie very close to one another, and the difference between the constant-$G$ and surface-dependent prescriptions remains comparatively small. The overall behavior is compact and smooth, indicating that, for the nearly spherical equilibrium shape, the surface character of the pairing interaction does not yet induce a major restructuring of the free-energy correction. In this regime, proton and neutron subsystems respond in a broadly similar manner, and the pairing contribution to the free energy remains moderate in magnitude.

The situation already changes in the second-saddle configuration, where the solutions begin to separate systematically for both protons and neutrons. The surface-dependent pairing force leads to a noticeably stronger negative free-energy correction at low temperatures, and this difference persists over a substantial thermal interval before vanishing as the critical region is approached. Thus, as with the gap itself, the superdeformed region marks the onset of a qualitatively new regime in which the pairing prescription begins to influence the free-energy landscape non-negligibly. In other words, the difference between $G={\rm const}$ and $G\propto surface$ is no longer a small quantitative detail, but begins to modify the microscopic free-energy balance along the fission path.

A particularly important result appears in the asymmetric scission configuration. Here, the constant-$G$ solutions for both protons and neutrons are already very close to zero over the low-energy part of the spectrum, which means that in this approximation the pairing contribution to the free energy is essentially quenched in the asymmetric pre-scission region. By contrast, when the surface dependence of the pairing strength is taken into account, the pairing free-energy correction remains clearly negative, reaching values of the order of $-3$~MeV at low excitation energies. This is a physically significant effect, especially for low-energy fission, because it shows that the surface-enhanced pairing interaction continues to provide a substantial lowering of the deformation free energy precisely in the region where the fragments are being formed. The asymmetric valley is therefore not only shaped by shell effects, but may also retain a non-negligible pairing stabilization if the interaction strength follows the increase of the effective nuclear surface.

The most dramatic effect is observed in the symmetric scission configuration. Once again, the constant-$G$ solutions are close to zero, which implies almost complete suppression of the pairing free-energy contribution in this approximation. For protons, the surface-dependent prescription produces a correction of roughly the same order as in the asymmetric valley, namely about $-3$~MeV at low excitation energy. For neutrons, however, the effect is much stronger: the pairing free-energy correction exceeds $6$~MeV in absolute value, i.e., it becomes more negative than $-6$~MeV at low excitation energies. This is a striking and potentially very important result. It demonstrates that in the symmetric valley, the surface dependence of the pairing interaction generates a very strong additional lowering of the free energy, especially in the neutron sector, whereas in the constant-$G$ approximation this contribution is almost entirely absent. The difference between the two prescriptions is therefore of the order of several MeV and directly affects the relative microscopic stability of symmetric near-scission configurations.

This finding may have consequences that go beyond the pairing sector itself. Symmetric fission is known to be associated, on average, with enhanced neutron emission compared with the asymmetric channel. It is therefore tempting to ask whether the strong negative pairing free-energy correction in the symmetric valley, especially for neutrons, may be correlated with the neutron-emission properties of this channel. A more negative pairing correction lowers the total deformation-free energy and may shift the corresponding scission-line configurations to lower microscopic energies. In such a situation, the balance between deformation, intrinsic excitation, and neutron-emission thresholds may be altered in a way that affects the neutron multiplicity. One may therefore speculate that the weaker pairing correction in the asymmetric valley is related to its comparatively lower neutron emission, whereas the much stronger neutron pairing contribution in the symmetric valley may be connected with the larger neutron yield characteristic of that channel. Although such a correlation cannot be established on the basis of the present figure alone, the result strongly suggests that the deformation dependence of the pairing free-energy correction should be considered as a potentially relevant ingredient in the microscopic interpretation of prompt-neutron emission systematics.

More generally, Fig.~\ref{fig:Fpair_all} shows that the pairing free-energy correction is even more sensitive to deformation and to the specific form of the pairing interaction than the gap itself. In the compact ground-state region, the two prescriptions remain relatively close, but as elongation increases, they diverge rapidly; in the scission region, they yield qualitatively different physical pictures. In the constant-$G$ approximation, the pairing free-energy correction becomes nearly irrelevant at large deformation, while in the surface-dependent scheme it remains sizable and, in the symmetric valley, even very large. This shows that the fate of pairing near scission cannot be inferred from ground-state intuition alone. On the contrary, once the surface of the fissioning nucleus grows and the neck develops, pairing may re-enter the problem in a much more consequential way through the free energy than one would conclude from compact shapes.
\begin{figure}[b]
\centering
\includegraphics[width=\columnwidth]{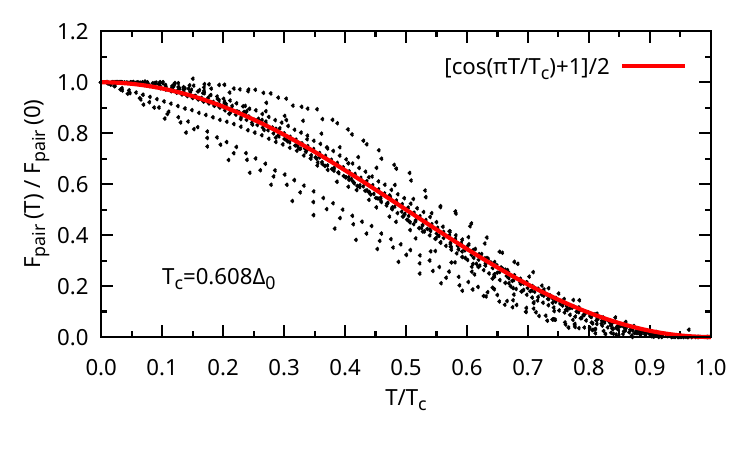}
\caption{Average temperature dependence of the normalized pairing free-energy correction. The black points represent the full set of calculated values, while the red curve corresponds to the approximation $F_{\rm pair}(T)/F_{\rm pair}(0)=[\cos(\pi T/T_c)+1]/2$ with $T_c=0.608\Delta_0$. 
}
\label{fig:Fpair_scaling}
\end{figure}

At the same time, the overall thermal behavior of the curves remains structurally similar to that observed in the gap itself: a finite negative correction at low temperatures, followed by a gradual reduction in magnitude and eventual disappearance as the critical region is approached. This suggests that, despite the very large differences in absolute scale between the various deformation points and pairing prescriptions, the temperature damping of the pairing free-energy correction is still governed by a common underlying mechanism. The main effect of deformation and of the surface dependence of the interaction is then to modify the characteristic scale of the correction, not to alter the generic form of its thermal attenuation. In this sense, Fig.~\ref{fig:Fpair_all} reinforces the conclusion that finite-temperature pairing along the fission path is characterized by a near-universal damping pattern, but with deformation-dependent amplitudes that become especially important in the scission region. Several additional features of Fig.~\ref{fig:Fpair_all} are worth emphasizing. First, unlike the gap itself, the quantity displayed here is directly the pairing contribution to the free energy, i.e., to the thermodynamic potential that enters the finite-temperature deformation landscape. The observed differences, therefore, do not merely reflect a formal change in the superfluid solution; they imply an actual modification of the local free-energy balance and may thus influence the competition between different fission valleys. In this sense, the figure shows directly where pairing remains an active component of the collective driving potential and where it becomes effectively irrelevant.

Second, the quenching of the pairing free-energy correction with increasing deformation is strongly dependent on the prescription. In the constant-$G$ approximation, the pairing contribution becomes very small already in the scission region, so that pairing is practically removed from the free-energy balance precisely where fragment formation takes place. By contrast, in the surface-dependent prescription, the pairing correction remains sizable and, in some cases, very large. This indicates that once the nuclear surface is strongly increased by elongation and neck formation, pairing may survive at the free-energy level far more efficiently than a constant monopole force would suggest. The result is therefore not only quantitative but also conceptual: the fate of pairing near scission depends qualitatively on whether the interaction is allowed to follow the changing geometry of the system.

Third, the proton and neutron sectors cease to behave in parallel as the system approaches scission. At the ground state and, to a lesser extent, at the second saddle, their behavior remains broadly similar. In the strongly deformed region, however, the neutron contribution becomes much more pronounced, especially in the symmetric valley. This suggests that deformation-induced restructuring of the single-particle spectrum affects neutron pairing more strongly than proton pairing near rupture. Since the neutron sector is also directly linked to prompt-neutron emission, this asymmetry may be of considerable physical relevance.

Fourth, the pairing free-energy correction remains negative up to temperatures close to the critical region and vanishes only gradually. This means that even when the pairing gap itself is already substantially reduced, the pairing field may still leave a visible imprint on the free-energy landscape. Pairing should therefore not be viewed as an abruptly disappearing ingredient, but rather as a correlation effect whose thermodynamic influence fades progressively with temperature. This observation supports the use of free-energy corrections, rather than gap values alone, when discussing the persistence of pairing near scission.

Finally, the contrast between the asymmetric and symmetric valleys appears especially significant. In the asymmetric channel, the surface-dependent prescription still yields a noticeable reduction in free energy, but the effect remains moderate. In the symmetric channel, and particularly for neutrons, the same mechanism produces a much larger correction. This points to a qualitative distinction between the two scission valleys: pairing seems to remain a secondary correction in the asymmetric case, whereas in the symmetric case it becomes a potentially major component of the microscopic free-energy balance. Since symmetric fission is also associated with enhanced neutron emission, this raises the intriguing possibility that the strong neutron pairing contribution in the symmetric valley is connected with the energy budget available for prompt-neutron release. Even if such a correlation cannot yet be demonstrated quantitatively, Fig.~\ref{fig:Fpair_all} strongly suggests that the deformation dependence of the pairing free-energy correction should be considered among the relevant microscopic ingredients in any discussion of scission energetics and neutron-emission systematics.

As shown in Fig.~\ref{fig:Fpair_scaling}, the calculated values collapse onto a simple average dependence. A particularly simple approximation to the proton ($p$) or neutron ($n$) pairing free-energy correction is obtained in the form: 
\begin{equation}
F_{\rm pair}^{p(n)}(T)\approx F_{\rm pair}^{p(n)}(0)\,
\frac{\cos\!\left(\pi T/T_c^{p(n)}\right)+1}{2}.
\label{eq:Fpair_scaling}
\end{equation}
where $T_c^{p(n)}$ is the effective critical temperature associated with the corresponding proton or neutron zero-temperature gap. Equation~(\ref{eq:Fpair_scaling}) is understood as a compact approximation to the microscopic pairing free-energy correction.

An important consequence of this result is that pairing correlations may still leave a visible trace in the free-energy landscape at deformation points close to scission, even when the corresponding gap is already strongly reduced. This observation will be relevant when discussing the odd-even effect in fragment charge distributions.

To use the calculated pairing free-energy correction in a macroscopic--microscopic framework, an additional subtraction is required. 
In this approach, the smooth macroscopic contribution to the nuclear energy is adjusted to experimental masses and therefore, in an average sense, already contains the pairing contribution.
As a consequence, the microscopic pairing term entering the temperature-dependent deformation energy should be understood not as the full pairing free energy itself, but rather as its fluctuating part with respect to the corresponding smooth average. This means that from the temperature-dependent pairing free-energy obtained microscopically one has to subtract the average pairing contribution, denoted by $\langle F_{\rm pair}(T)\rangle$.

The temperature dependence of this average quantity may be estimated by combining the scaling relation of Eq.~(\ref{eq:Fpair_scaling}) with the corresponding critical-temperature parametrization of Eq.~(\ref{eq:Delta_scaling_params}). To evaluate the average pairing contribution itself, we use the so-called uniform approximation~\cite{Brack1972}, in which the single-particle spectrum is characterized by a smooth level density. In this approximation, there is a one-to-one relation between the average zero-temperature pairing gap, denoted by $\tilde\Delta_0$, and the pairing-strength parameter $G$:
\begin{equation}
\frac{2}{G}=2\tilde g(\tilde\lambda)\ln\left(\frac{2\Omega}{\tilde\Delta_0}\right),
\qquad {\rm or} \qquad
\tilde\Delta_0=2\Omega\, \exp{\Big(-\frac{1}{\tilde g(\tilde\lambda)\,G}\Big)} ~.
\label{eq:avdelta}
\end{equation}
Here, $\tilde g(\tilde\lambda)$ is the smooth single-particle level density at the average Fermi energy $\tilde\lambda$, while $2\Omega$ denotes the width of the pairing window. If $n_c$ single-particle levels are included on each side of the Fermi energy, so that $2n_c$ levels are taken in total, the pairing window width reads
\begin{equation}
2\Omega=\frac{2n_c}{\tilde g(\tilde\lambda)}~.
\label{eq:pairing_window}
\end{equation}

Within the same approximation, the average zero-temperature pairing contribution to the free energy may be estimated as
\begin{equation}
\langle F_{\rm pair}(0)\rangle = -\frac{1}{2}\tilde g(\tilde\lambda)\tilde\Delta_0^2 ~.
\label{eq:avEpair}
\end{equation}
The temperature dependence of the average pairing free energy is then approximated by using the same damping law as for the microscopic pairing free-energy correction,
\begin{equation}
\langle F_{\rm pair}(T)\rangle
=\langle F_{\rm pair}(0)\rangle\,
\frac{\cos(\pi T/\tilde T_c)+1}{2}~,
\label{eq:avFpair}
\end{equation}
where $\tilde T_c$ is the effective critical temperature corresponding to the average gap $\tilde\Delta_0$. According to Eq.~(\ref{eq:Delta_scaling_params}), it is given by
\begin{equation}
\tilde T_c = 0.608\,\tilde\Delta_0~.
\label{eq:avTc}
\end{equation}

The final expression for the total temperature-dependent pairing free-energy correction is therefore written as
\begin{equation}
\delta F_{\rm pair}(T)=
F^p_{\rm pair}(T)-\langle F^p_{\rm pair}(T)\rangle
+F^n_{\rm pair}(T)-\langle F^n_{\rm pair}(T)\rangle ~,
\label{eq:dFpair}
\end{equation}
where $F^p_{\rm pair}(T)$ and $F^n_{\rm pair}(T)$ denote the microscopic proton and neutron pairing free-energy corrections, respectively, while $\langle F^p_{\rm pair}(T)\rangle$ and $\langle F^n_{\rm pair}(T)\rangle$ are their corresponding smooth average parts. The quantity $\delta F_{\rm pair}(T)$ thus represents the fluctuating pairing contribution that should be added to the smooth macroscopic energy. Together with the analogous shell correction discussed in the next section, it determines the temperature-dependent microscopic contribution to the potential-energy surface.


\section{Thermal damping of the shell free-energy correction}
\label{sec:shell_damping}
To complete the microscopic picture near scission, it is not sufficient to analyze only the thermal quenching of pairing. One must also consider the shell effect itself. In particular, it is important to distinguish between the shell correction associated with the underlying single-particle spectrum and the effect related to the survival of pairing correlations near the scission line. We therefore turn now to the question of how rapidly the shell correction is attenuated with increasing excitation energy along the scission line.
The shell correction is evaluated using the Strutinsky averaging procedure generalized to finite temperature. The average level density is written as
\begin{equation}
\widetilde g(e)=\frac{2}{\gamma}\sum_{\nu} f\left(\frac{e_{\nu}-e}{\gamma}\right),
\label{eq:gtilde}
\end{equation}
with
\begin{equation}
f(x)=\frac{\exp(-x^2)}{\sqrt{\pi}}\sum_{n=0,2,\dots}^{M} a_n H_n(x),
\label{eq:strutinsky_kernel}
\end{equation}
where $\gamma$ is the smoothing width and $H_n(x)$ are Hermite polynomials.

For a system of independent particles at temperature $T$, the energy is
\begin{equation}
E(T)=2\sum_{\nu} e_{\nu} n(e_{\nu}),
\label{eq:E_independent}
\end{equation}
where
\begin{equation}
n(e)=\frac{1}{1+\exp[(e-\widetilde\lambda)/T]}
\label{eq:Fermi_Dirac_avg}
\end{equation}
is the Fermi--Dirac occupation probability, and $\widetilde\lambda$ is determined from the particle-number condition. The corresponding Strutinsky-averaged energy and entropy are
\begin{align}
\widetilde E(T) &= \int_{-\infty}^{\infty} e\,\widetilde g(e)n(e)\,de,
\label{eq:Etilde}\\
\widetilde S(T) &= -\int_{-\infty}^{\infty} \widetilde g(e)\left[n(e)\ln n(e)
+\left(1-n(e)\right)\ln\left(1-n(e)\right)\right]de.
\label{eq:Stilde}
\end{align}
The smoothed Fermi energy is fixed by the particle-number condition,
\begin{equation}
N=\int_{-\infty}^{\infty}\widetilde g(e)n(e)\,de.
\label{eq:number_tilde}
\end{equation}
The shell correction to the energy and to the free energy are then defined as
\begin{align}
\delta E_{\rm shell}(T) &= E(T)-\widetilde E(T),
\label{eq:Eshell}\\
\delta F_{\rm shell}(T) &= \left[E(T)-T S(T)\right]
-\left[\widetilde E(T)-T\widetilde S(T)\right].
\label{eq:Fshell}
\end{align}

\begin{figure*}[t]
\centering
\hspace*{-0.5cm}%
\includegraphics[width=0.53\textwidth]{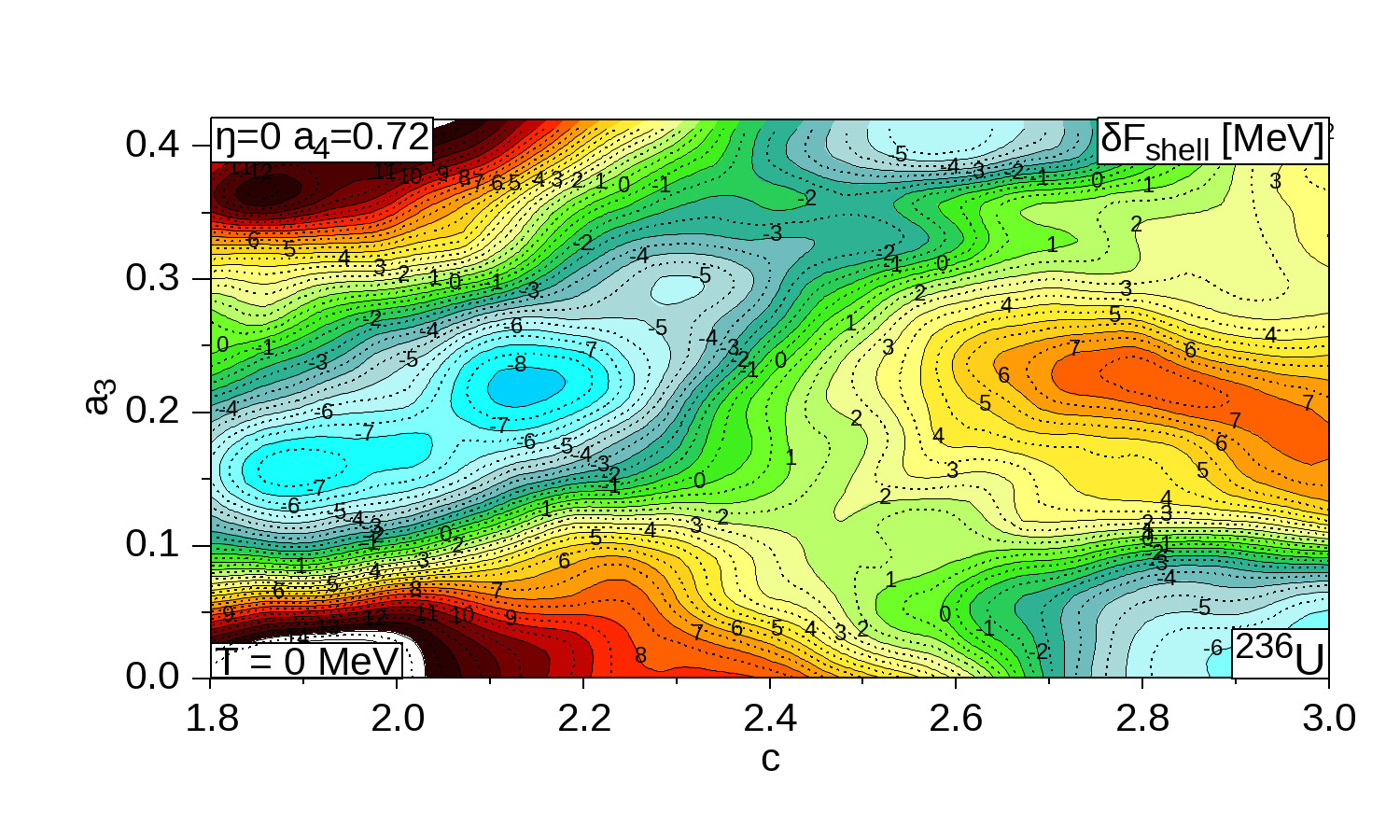}
\hspace{-0.055\textwidth}%
\includegraphics[width=0.53\textwidth]{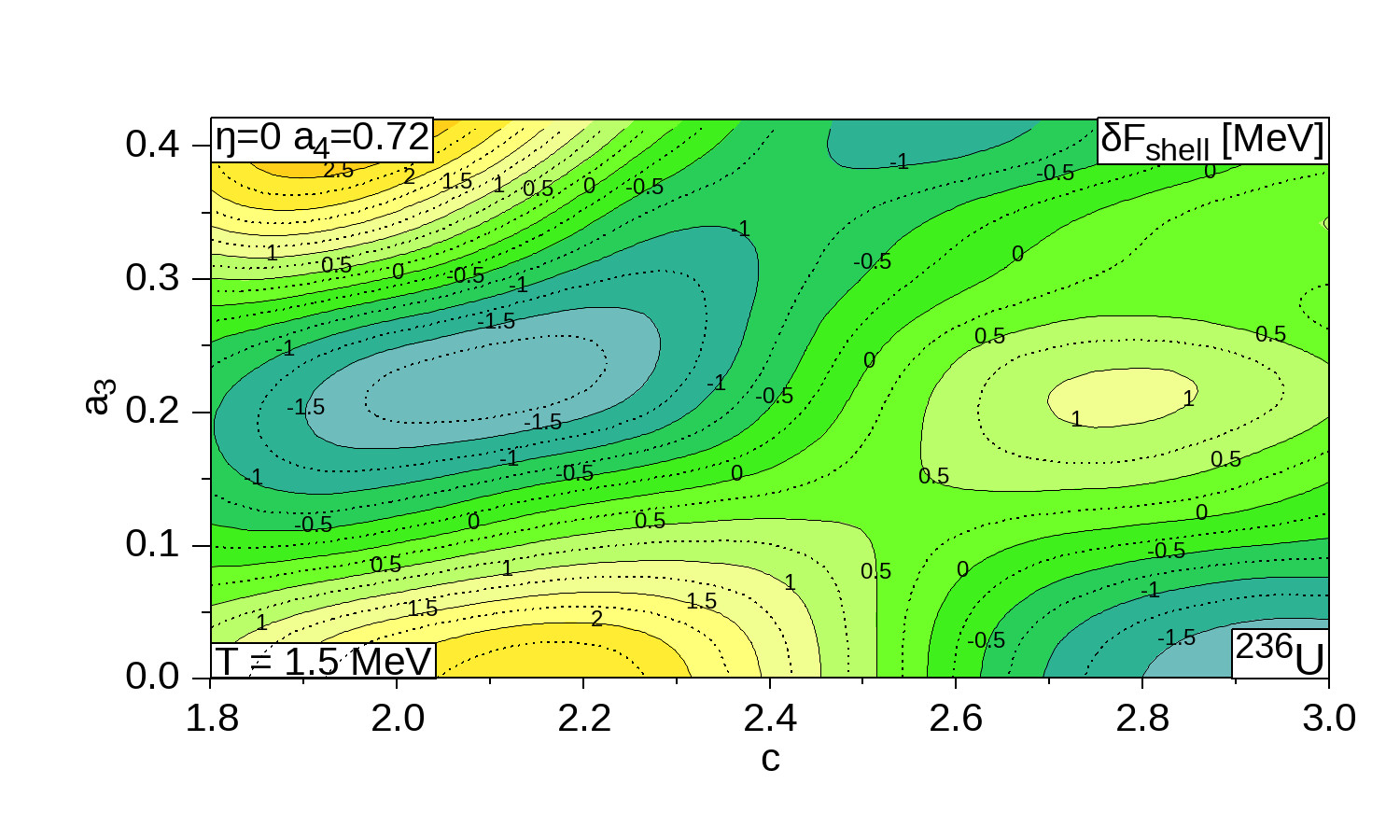}\\[0.2ex]
\hspace*{-0.5cm}%
\includegraphics[width=0.53\textwidth]{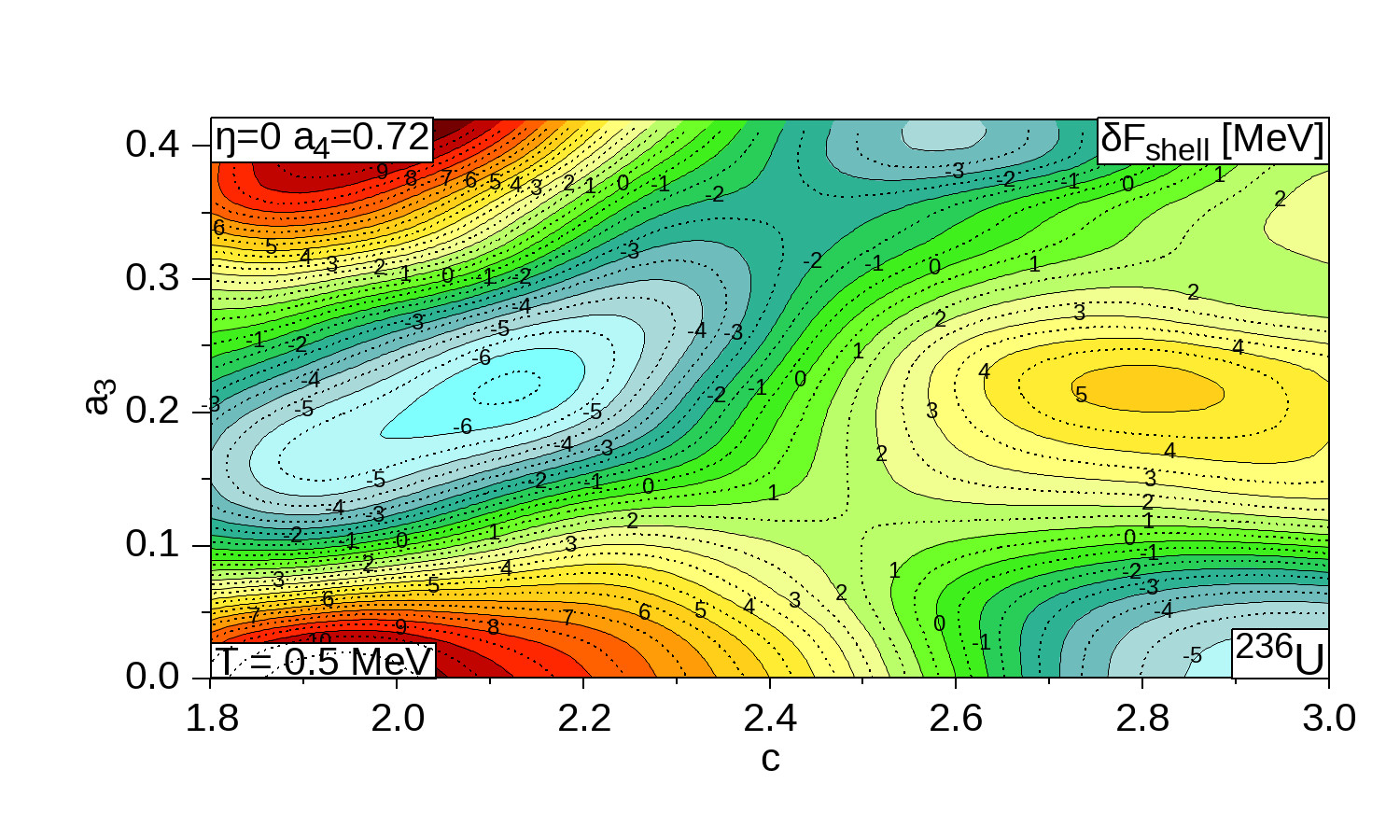}
\hspace{-0.055\textwidth}%
\includegraphics[width=0.53\textwidth]{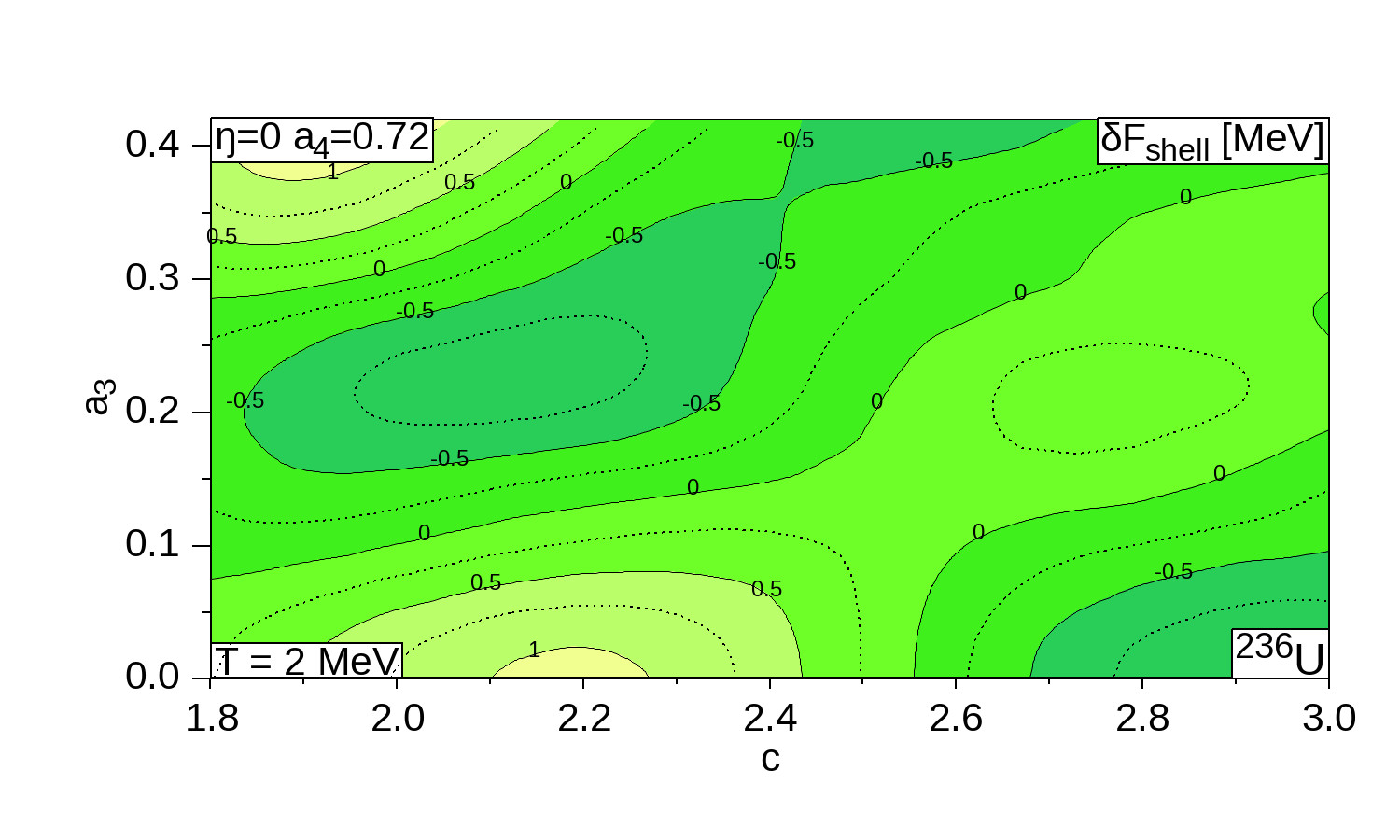}\\[0.2ex]
\hspace*{-0.5cm}%
\includegraphics[width=0.53\textwidth]{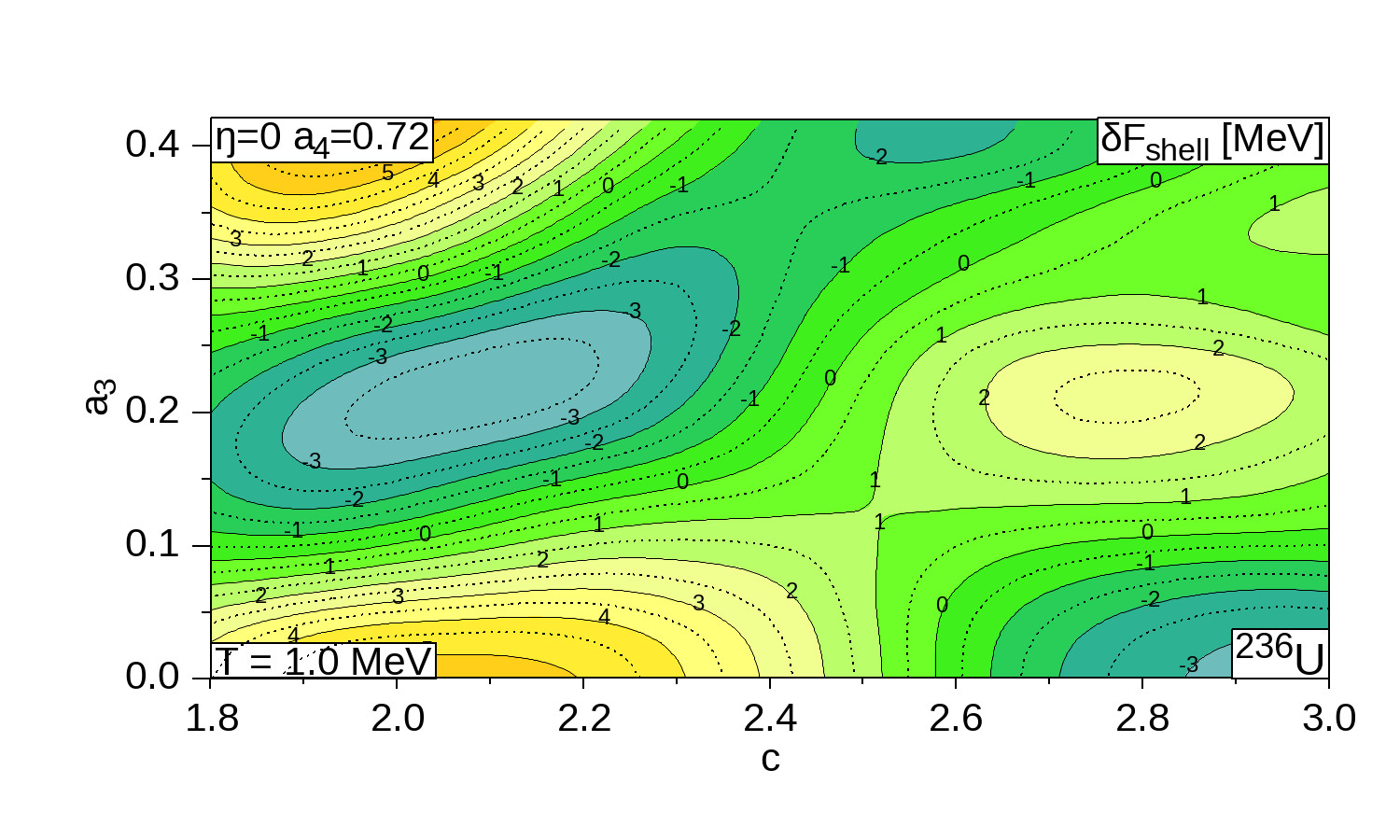}
\hspace{-0.055\textwidth}%
\includegraphics[width=0.53\textwidth]{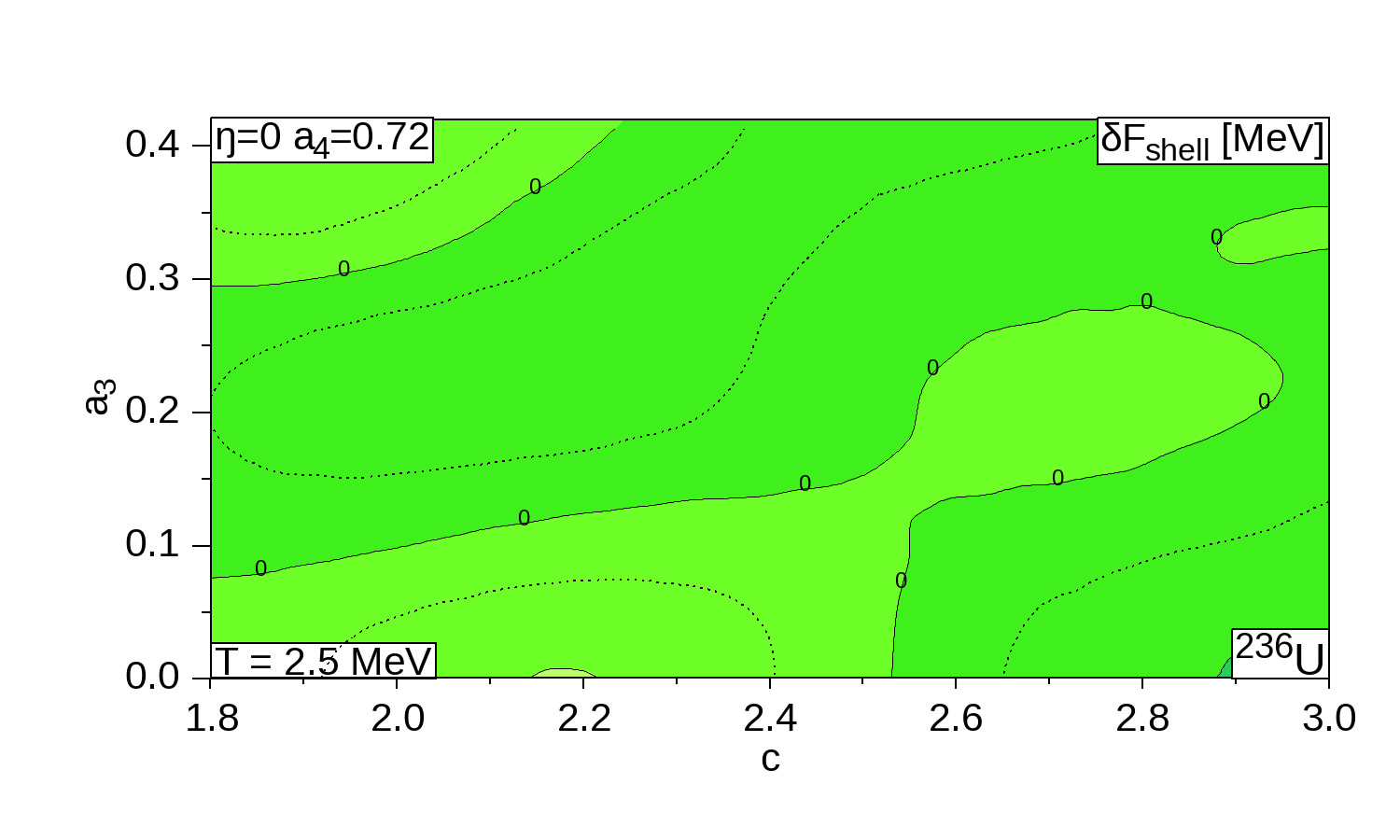}
\caption{Temperature evolution of the shell free-energy correction around scission for $^{236}$U in the $(c,a_3)$ plane at fixed $a_4=0.72$. The left column corresponds to $T=0$, $0.5$, and $1.0$ MeV, while the right column shows $T=1.5$, $2.0$, and $2.5$ MeV. With increasing temperature, the shell-induced structure is gradually washed out, and above about $2$--$2.5$ MeV the free-energy landscape becomes nearly flat.}
\label{fig:scission_maps_allT}
\end{figure*}

Figure~\ref{fig:scission_maps_allT} provides a particularly transparent picture of how the shell correction to the free energy evolves with temperature along the scission line defined by the axial FoS parametrization with fixed neck parameter $a_4=0.72$. The maps are shown in the $(c,a_3)$ plane, where $c$ denotes elongation and $a_3$ the mass-asymmetry degree of freedom.

At $T=0$, the shell structure is very pronounced and strongly corrugated. Several well-developed minima and maxima are visible, indicating that the shell contribution remains a dominant component of the deformation-free energy even for these strongly elongated pre-scission shapes. The deepest minimum is located around $a_3\approx 0.2$ and $c\approx 2.1$, where the shell correction reaches values of the order of $-8$~MeV. This is a remarkably strong effect, comparable to the largest shell corrections usually associated with compact equilibrium configurations. In addition to this particularly deep pocket, one can identify another favorable region for somewhat more compact and less asymmetric shapes, as well as a distinct minimum at larger elongation and stronger asymmetry, around $a_3\approx 0.4$ and $c\approx 2.6$--$2.8$. A further minimum is also seen in the more nearly symmetric division region. Thus, even at zero temperature, the map shows that shell effects along the scission line are not a small residual modulation but a major factor shaping the free-energy landscape and, therefore, the relative preference for different fragment partitions. In particular, the presence of both asymmetric and near-symmetric shell-favored regions indicates that the shell effect may be directly involved in the competition between different fission channels and in the partition of shell stabilization between the emerging fragments.

At $T=0.5$~MeV, the first signs of thermal damping are already clearly visible. The shell landscape remains recognizable, but its fine structure begins to soften. The previously more structured left-hand minimum becomes smoother and less sharply resolved; what looked at $T=0$ like a more internally differentiated shell pocket now tends to merge into a broader depression. At the same time, the minimum becomes visibly shallower, and the high positive shell regions, pronounced at zero temperature, are systematically reduced in magnitude. Thus, the first action of temperature is not simply to lower the entire map uniformly, but to weaken its contrast: both the attractive shell pockets and the repulsive shell ridges are attenuated, so that the entire landscape becomes less rugged. Even such a modest temperature increase is sufficient to begin to wash out the sharp shell selectivity present at $T=0$.

At $T=1.0$~MeV, this trend becomes even more pronounced. The major shell-favored regions can still be identified, but they are already much broader, more diffuse, and significantly less deep than at lower temperatures. The main minimum around $(c,a_3)\approx(2.1,0.2)$ is still visible, but its internal structure has essentially disappeared, and the same applies to the more asymmetric minimum at larger elongation. What remains is the map's large-scale topology rather than its detailed oscillatory content. This is an important point: the first component of the shell correction to disappear is the fine local structure --- the sharp pockets, narrow ridges, and small-scale oscillations --- whereas the broader organization of the landscape survives somewhat longer. In other words, temperature first destroys the detailed shell selectivity of the scission line, but not yet its overall tendency to favor certain deformation regions over others.

At $T=1.5$~MeV, the shell structure is further attenuated. The minima survive only as broad and shallow depressions, and the map is now dominated by long-wavelength variations rather than sharply localized shell pockets. The shell correction remains nonzero and introduces some modulation into the free-energy surface, but it no longer acts as a strongly selective mechanism. Instead, it provides only a broad bias favoring some regions of deformation space over others. The contrast between competing valleys is much weaker, and the shell effect can no longer enforce narrow-channel selection as it could at low temperatures.

By $T=2.0$~MeV, the shell structure is already very close to complete disappearance. Only weak remnants of the earlier minima are still discernible, and even these survive merely as broad, shallow depressions. Both the asymmetric and the more nearly symmetric shell-favored regions remain faintly visible, which is itself instructive: thermal damping does not eliminate one of these channels much earlier than the other, but gradually reduces the shell-driven distinction between them. This implies that there exists an intermediate thermal regime in which shell effects no longer impose sharply defined valleys, yet still bias the population of competing scission configurations. Such a regime may be particularly important dynamically because the system may still remember the shell landscape, but in a broader, less deterministic way.

Finally, at $T=2.5$~MeV, the shell-induced structure is practically gone. The free-energy correction becomes nearly featureless over the whole $(c,a_3)$ plane, and only very weak residual modulations remain. The oscillatory character of the shell correction has essentially disappeared, and the map is almost flat. At this stage, shell effects can no longer be expected to play a substantial role in determining the preferred scission configuration. The competition between different deformation channels must then be predominantly governed by the macroscopic component of the free energy, together with dynamical effects, rather than by localized shell stabilization.

Several broader conclusions emerge from this sequence. First, the shell correction at scission can be very strong at low temperature, with amplitudes comparable to those familiar from compact ground-state shapes. This shows that the shell effect remains a major microscopic ingredient even along the scission line and may therefore strongly influence the final partitioning of fragments. Second, the temperature dependence of damping is continuous and highly structured rather than abrupt. The evolution proceeds in a clear hierarchy: first, the fine oscillatory shell structure is smoothed out; then the depths of the major minima are reduced; then the contrast between competing valleys is weakened; and only at the highest temperatures does the entire landscape become nearly flat. This ordered progression is physically meaningful because it shows that the shell effect passes through an intermediate regime in which its detailed pattern is destroyed, while its large-scale influence persists. Third, the global topology of the map is more robust than the local details: the main asymmetric and near-symmetric favorable regions remain identifiable over the first few temperature steps even after their fine substructure has disappeared. Thus, the long-wavelength component of the shell correction is more resistant to thermal smearing than the short-wavelength oscillatory component.

Another point worth emphasizing is that the thermal evolution is not completely uniform over the plane. Some regions flatten faster than others. The left-hand shell minimum at smaller elongation loses depth and internal structure very early, whereas the right-hand minimum at larger elongation remains recognizable somewhat longer as a broad feature. This suggests that different regions of the scission line are characterized by distinct effective shell spacings and, therefore, by different resistance to thermal averaging. In this sense, the damping of shell effects is deformation dependent not only in amplitude but also in spatial pattern.

\begin{figure}[b]
\centering
\includegraphics[width=\columnwidth]{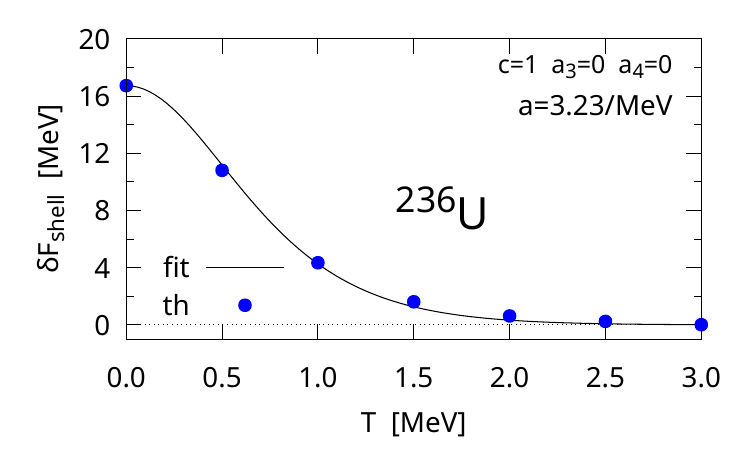}
\includegraphics[width=\columnwidth]{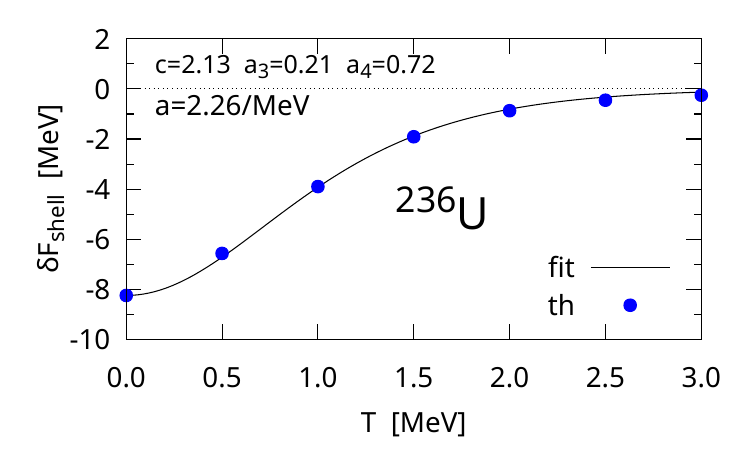}
\caption{Average temperature dependence of the normalized shell free-energy correction of spherical (top) nucleus $^{236}$U and near the asymmetric scission configuration (bottom). The blue points correspond to the calculated values obtained with the Yukawa-folded single-particle potential, while the red curve shows the present fit. The damping is well reproduced by $F_{\rm shell}(T)/F_{\rm shell}(0)=aT/\sinh(aT)$ with $a\approx 3.23~\mathrm{MeV}^{-1}$ for the spherical shape and $a\approx 2.26~\mathrm{MeV}^{-1}$ at scission.}
\label{fig:Fshell_fit}
\end{figure}

Figure~\ref{fig:Fshell_fit} shows the average temperature dependence of the shell correction to the free energy for the spherical shape (top) and in the vicinity of the asymmetric scission configuration (bottom) of $^{236}$U. The calculated points are reproduced remarkably well by the simple attenuation law
\begin{equation}
\delta F_{\rm shell}(T)=\delta F_{\rm shell}(0)\,\frac{aT}{\sinh(aT)}~,
\label{eq:Fshell_BM_like}
\end{equation}
with a fitted coefficient $a\approx 3.23\,{\rm MeV^{-1}}$ for the spherical shape and $a\approx 2.26\,{\rm MeV^{-1}}$ around the asymmetric scission point. It is worth emphasizing that this form is closely related to the classical Bohr--Mottelson prescription for the temperature damping of shell effects~(\cite[p.~608]{BMbook}). The quality of the fit is striking: over the entire temperature range considered, from $T=0$ to about $3$~MeV, the analytical curve closely follows the calculated shell free-energy correction. This demonstrates that the thermal damping of the shell contribution, although extracted here from a fully microscopic calculation based on the Yukawa-folded single-particle potential, can be represented to a very good approximation by a compact one-parameter formula. A useful way to interpret the fitted damping coefficient is to combine it with the characteristic harmonic-oscillator energy scale, thereby defining an effective dimensionless constant.

In fact, the present results suggest that the Bohr--Mottelson-type scaling holds not only near spherical shapes, where it is usually introduced, but also at strongly deformed configurations, including the asymmetric scission region. The fitted values obtained for the spherical shape and for the asymmetric scission point are consistent with one another in a physically meaningful way: the ratio of the extracted damping parameters is approximately inverse to the corresponding ratio of the effective Harmonic Oscillator (H.O.) $\hbar\omega_0$ values characterizing the local single-particle spectrum. This indicates that the deformation dependence of the thermal damping arises primarily from that of the characteristic shell spacing.

In the Bohr and Mottelson model of the free-energy temperature dependence, the damping parameter $a_{\rm BM}$ depends on the average distance between H.O. shells:
\begin{equation}
a_{\rm BM}=\frac{2\pi^2}{\hbar\omega_0}~.
\label{eq:BMa}
\end{equation}
Assuming $\hbar\omega_0({\rm sph.})=\hbar\stackrel{\circ}{\omega}_0=40/A^{1/3}$ one obtains $a_{\rm BM}$= 3.05 MeV$^{-1}$ for the spherical $^{236}$U nucleus. This value is very close to the one fitted above ($a$=3.23 MeV$^{-1}$) for the spherical shape. It is well known that the H.O. frequency $\omega_0({\rm def.})$ grows with the deformation of the nucleus (conf. e.g. \cite{Nilsson1969}), e.g., for the axially symmetric spheroid with the half-axis $c$ one has:
\begin{equation}
{\omega_0(c)\over{\stackrel{\circ}{\omega}_0}}=\left({2\sqrt{c}+1/c\over 3}\right)^3~.
\label{eq:hw0}
\end{equation}
So, according to Eq.~(\ref{eq:BMa}), the damping parameter $a$ should decrease with the elongation of the nucleus ($2c$). It aligns with the results of our fits. Namely, the ratio $a({\rm sph.})/a({\rm sciss.})=3.23/2.26\simeq 1.44$ is nearly equal to the inverse ratio of the corresponding H.O. frequencies ${\omega_0(2.13)\over{\stackrel{\circ}{\omega}_0}}\simeq 1.45$. This result validates the Bohr--Mottelson prescription well.

This result strongly suggests that the thermal damping of the shell free-energy correction follows an approximately universal scaling law, with the relevant temperature variable normalized to the local shell spacing. From this point of view, deformation does not alter the form of the damping law itself but mainly changes the local single-particle energy scale that enters the normalization. The attenuation of shell effects thus acquires a natural geometrical interpretation, i.e., it only depends on the deformation of the nucleus:
\begin{equation}
a=\frac{b}{\hbar\omega_0({\rm def.})}~~~~{\rm with}~~~b\approx 21~.
\label{eq:dump-b}
\end{equation}
Note, the fitted value of the parameter $b$ is very close to the Bohr--Mottelson constant in Eq.~(\ref{eq:BMa}) $2\pi^2=19.74~$. As the shape evolves, the local shell spacing changes, and with it, the temperature scale over which the shell structure is smoothed out.

This outcome is physically instructive for several reasons. First, it confirms that the shell correction to the free energy decays smoothly and systematically with temperature, rather than disappearing abruptly. Second, it shows that this damping may be described by a scaling law that holds over a broad range of deformations, including shapes far from sphericity. Third, it suggests that the deformation dependence of the shell damping can, to a large extent, be absorbed into that of the local oscillator scale. In this sense, the present fit provides a compact, physically transparent parametrization of the shell free-energy correction, suitable for practical implementation in finite-temperature fission calculations. It is also encouraging that the same scaling law is expected to work in lighter nuclei, owing to the explicit $A$ dependence of the oscillator constant. A similar conclusion has recently been reached by Ivanyuk \textit{et al.}~\cite{Ivanyuk2025} in their five-dimensional Langevin study of shell effects and multichance fission in the sub-lead region. They showed that the microscopic free-energy correction decreases with excitation energy more slowly than predicted by the standard Ignatyuk prescription. Their results therefore support the view that shell effects may persist up to relatively high excitation energies and should be described within a finite-temperature free-energy framework rather than by a simple exponential damping law.

\section{Stochastic Fission Dynamics in the Fourier-over-Spheroid Framework}
\label{sec:method}

To determine the fragment partition and the associated isotope-resolved observables, we employ a stochastic description of fission dynamics formulated within the four-dimensional Langevin Equation framework. This approach provides a unified treatment of collective transport from saddle to scission, dissipative conversion of collective energy into intrinsic excitation, pre-scission neutron evaporation including multi-chance fission, and correlated post-scission de-excitation of the nascent fragments~\cite{Pomorski2024LM}. It is therefore particularly well suited for the present purpose, where the aim is not only to describe the average fission path but also to generate an event ensemble from which fragment mass, charge, kinetic energy, and neutron-emission observables can be extracted consistently.

The collective evolution is propagated through the multidimensional Langevin equations written for the generalized collective coordinates $\vec q=\{q_i\}$ and their conjugate momenta $\vec p=\{p_i\}$ as
\begin{equation}
\dot q_i=\sum_{j}\big[{\cal M}^{-1}(\vec q\,)\big]_{ij}\,p_j ,
\label{eq:langevin_q}
\end{equation}
\begin{equation}
\begin{aligned}
\dot p_i={}&-\frac{1}{2}\sum_{j,k}
\frac{\partial\big[{\cal M}^{-1}(\vec q\,)\big]_{jk}}{\partial q_i}\,p_jp_k
-\frac{\partial V(\vec q\,)}{\partial q_i} \\
&-\sum_{j,k}\gamma_{ij}(\vec q\,)\big[{\cal M}^{-1}(\vec q\,)\big]_{jk}\,p_k
+{\cal F}_i(t)\,,
\end{aligned}
\label{eq:langevin_p}
\end{equation}
where ${\cal M}(\vec q\,)$ is the collective inertia tensor, $\gamma(\vec q\,)$ the friction tensor, $V(\vec q\,)$ the thermodynamic driving potential, and ${\cal F}_i(t)$ the stochastic force. In the present implementation, the inertia tensor is evaluated in the irrotational-flow approximation, while dissipation is described by wall-type friction, following Ref.~\cite{Bartel2019CPC}. Fluctuations are coupled to dissipation in the standard Langevin manner, including the effective-temperature correction accounting for quantum fluctuations~\cite{PomorskiHofmann1981}. The resulting stochastic differential equations are solved numerically to generate an ensemble of saddle-to-scission trajectories; each terminating trajectory defines one scission event from which fragment observables are constructed.
The driving potential entering the Langevin equations is constructed from a microscopic--macroscopic energy functional evaluated for FoS shapes. The underlying shape parametrization, together with the reference configurations used in the present work, was introduced in Sec.~\ref{sec:shape}~\cite{Pomorski2023FoS,PomorskiNerlo2023FoSNonax}.
The macroscopic part is taken either from the Lublin--Strasbourg Drop (LSD) model~\cite{PomorskiDudek2003LSD} or from the iso-scalar liquid-drop approximation (ISOLDA)~\cite{PomorskiXiao2025ISOLDA}, depending on the variant considered. The microscopic shell and pairing corrections are obtained from a Yukawa-folded single-particle potential~\cite{Dobrowolski2016YukawaFolded}. Since dissipative transport from saddle to scission generates intrinsic excitation, the collective drift must be formulated in thermodynamic rather than purely static terms. For this reason, the Langevin propagation is driven by the Helmholtz free energy, constructed from temperature-dependent macroscopic and microscopic contributions, as well as the deformation-dependent level-density parameter. The quadratic temperature growth of the macroscopic sector and the damping of microscopic corrections with increasing temperature are introduced consistently within the established  prescription~\cite{NerloPomorska2006PRC}. In this way, the collective dynamics is governed by a deformation- and temperature-dependent free-energy landscape rather than by the zero-temperature potential-energy surface alone.

The pre-scission competition between fission and neutron evaporation is described by a coupled set of Langevin equations, following the strategy introduced previously in Refs.~\cite{Pomorski2000NPA,Pomorski2024LM}, but here implemented within the FoS shape representation. After scission, the primary fragments undergo statistical de-excitation, treated event-by-event, which preserves correlations among fragment mass, charge, kinetic energy, and neutron emission~\cite{Pomorski2024LM}. In the excitation-energy range relevant to the present study, charged-particle emission is strongly suppressed, and neutron evaporation dominates. The neutron-emission widths are calculated within the Weisskopf--Ewing formalism~\cite{Delagrange1986}, using the Dostrovsky--Fraenkel--Friedlander parametrization for inverse cross sections~\cite{Dostrovsky1959} and standard Fermi-gas level densities with deformation-dependent systematics taken from Ref.~\cite{NerloPomorska2002PRC}. In this way, the entire   chain from dissipative stochastic transport in the FoS collective space to correlated statistical de-excitation --- provides the basis for calculating fragment partitions and the isotope-resolved observables analyzed in the present work.


\section{Charge polarization and odd--even effect at scission}
\label{sec:cheq}

The present interpretation of the origin of odd--even effects in fission-fragment charge yields connects naturally with our earlier Langevin-based treatment of charge partition at scission, in which the most probable fragment charge was determined by minimizing the scission energy, while neighboring integer partitions were populated according to a Wigner-type probability prescription~\cite{Pomorski2024LM}. Within that framework, odd--even staggering could already be generated at the level of the final charge distribution, but its temperature dependence entered only through an effective prescription and therefore lacked a direct microscopic foundation.

From the perspective developed in the present work, the crucial point is that if the pairing free-energy correction remains non-negligible in the pre-scission and scission region, then the discrete charge partition established at neck rupture should still retain a memory of residual pairing correlations. In that case, the odd--even effect observed in the final fragment charge yields may be regarded as a genuine observable signature of pairing survival at large deformation. Conversely, once the pairing contribution is sufficiently attenuated with increasing temperature, the local charge-partition landscape becomes smoother, the energetic distinction between neighboring even and odd charge splits is reduced, and the odd--even staggering should gradually weaken or disappear.

This interpretation is fully consistent with our earlier isotope-resolved Ba and Xe calculations, where the calculated charge distributions were typically smoother than the evaluated data precisely in those cases where experimental odd--even staggering remained visible~\cite{Pomorski2024LM}. That discrepancy pointed to a missing or overly suppressed pairing-related structure in the model description. The present finite-temperature pairing formalism provides the microscopic basis for improving this situation by replacing the previous effective damping of the total microscopic contribution with separate temperature- and deformation-dependent pairing and shell terms. In this sense, the present analysis does not merely refine the phenomenology of charge polarization, but identifies the specific microscopic ingredient that controls the persistence of the fine charge structure near scission.

This point is further supported by the global systematics summarized in Table~\ref{tab:odd_even_statistics}. The table shows that the occurrence of odd--even staggering is not random across the actinide region, but follows a clear energetic pattern: the effect survives only when the total energy available in the late stage of the process remains below a characteristic upper limit. Although the precise numerical threshold depends somewhat on the underlying macroscopic prescription, the general trend is robust and strongly suggests that the visibility of odd--even staggering is governed by the competition between intrinsic excitation and residual pairing correlations in the scission region. The systematic weakening of the effect for heavier systems and for reactions with larger initial excitation is therefore naturally consistent with the thermal damping mechanism discussed above.

\begin{figure}[t]
\centering
\includegraphics[width=0.48\columnwidth]{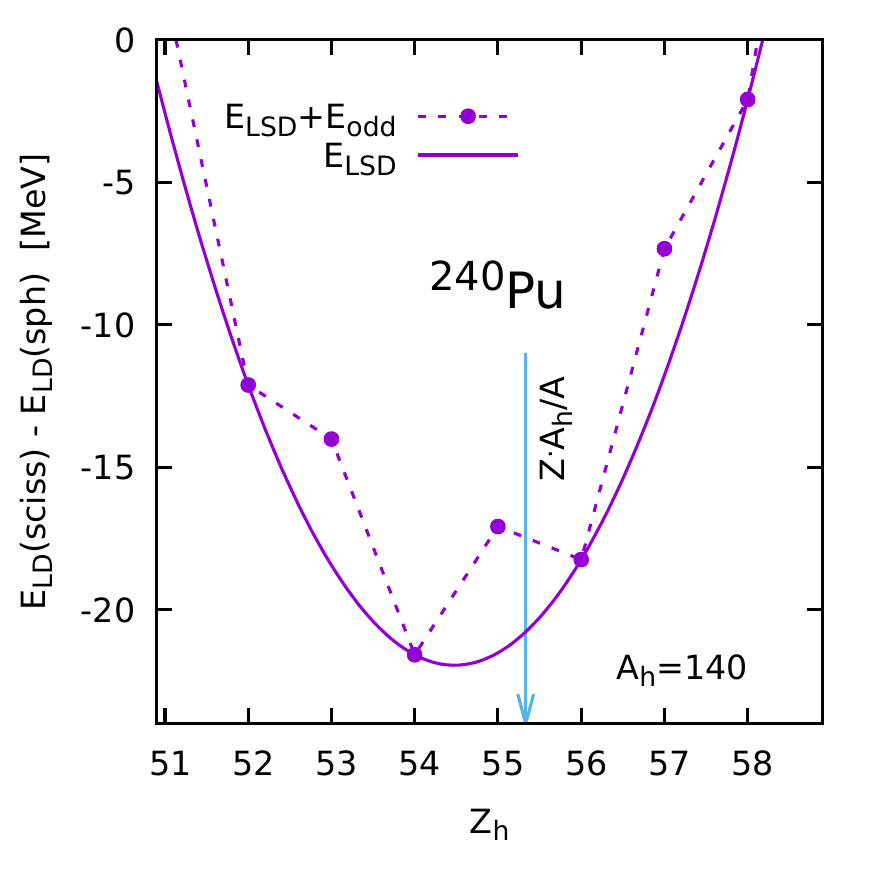}
\includegraphics[width=0.48\columnwidth]{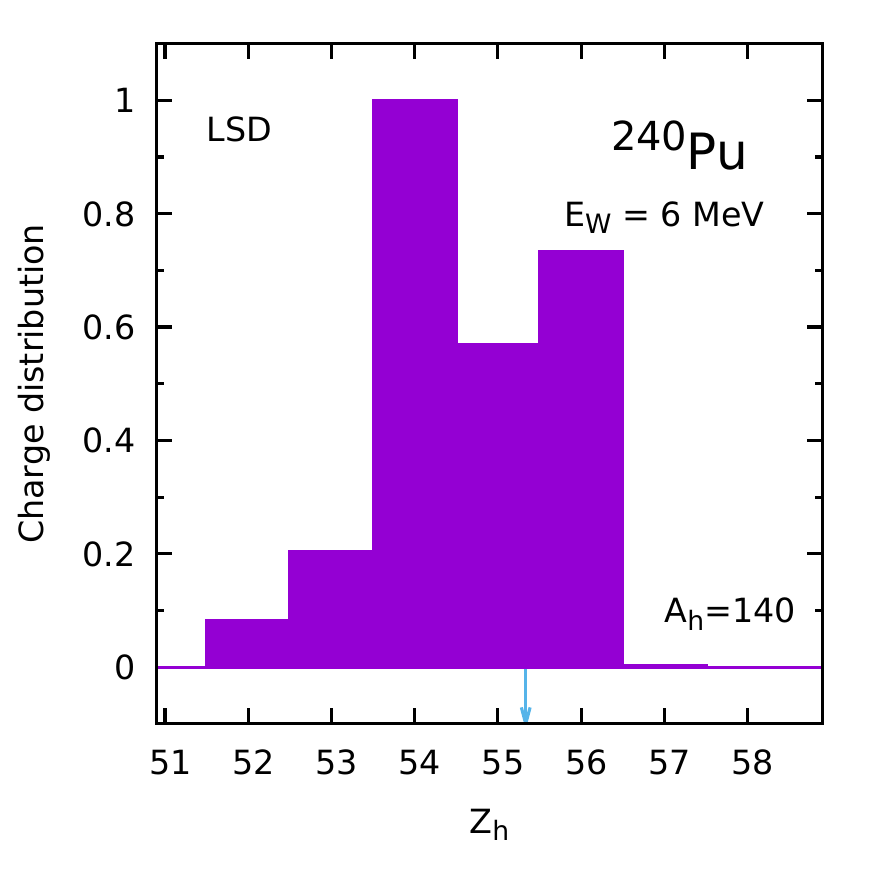}
\caption{Illustration of the charge-polarization mechanism and the resulting discrete charge distribution at scission for a representative split. Left: scission energy as a function of the heavy-fragment charge, calculated within the LSD-based macroscopic description with and without the odd--even term. Right: corresponding charge distribution obtained from the Wigner-type probability prescription. These plots show how a residual pairing-related odd--even contribution may bias the final integer charge partition.}
\label{fig:charge_odd_even_scheme}
\end{figure}

To describe the charge partition at scission, let $Z_h$ and $A_h$ denote the proton and mass numbers of the heavy fragment. The corresponding light-fragment values are $
Z_l=Z-Z_h,A_l=A-A_h,$
where $Z$ and $A$ are the charge and mass numbers of the fissioning nucleus. The vectors $\vec q_h$ and $\vec q_l$ denote the collective deformation coordinates assigned to the heavy and light prefragments at scission, while $R_{12}$ is the distance between their charge centers. The scission energy may then be written as
\begin{equation}
\begin{aligned}
V_{\rm sc}(Z_h)={}&
E_{\rm LSD}(Z_h,A_h;\vec q_h)
+E_{\rm LSD}(Z_l,A_l;\vec q_l) \\
&+\frac{e^2 Z_h Z_l}{R_{12}}
- E_{\rm LSD}(Z,A;0) .
\end{aligned}
\label{eq:Vsc_Zh}
\end{equation}
Here, the first two terms represent the macroscopic energies of the nascent fragments, the third term is their Coulomb interaction energy, and the last term subtracts the macroscopic energy of the initial spherical compound nucleus. Strictly speaking, $V_{\rm sc}$ depends not only on $Z_h$, but also parametrically on $\vec q_h$, $\vec q_l$, and $R_{12}$.

The corresponding charge distribution is modeled by a Wigner-type expression,
\begin{equation}
W(Z_h)=\exp\left\{
-\frac{\left[V_{\rm sc}(Z_h)-V_{\rm sc}(Z_{\min})\right]^2}{E_0^2}
\right\},
\label{eq:Wigner_charge}
\end{equation}
where $Z_{\min}$ is the value of $Z_h$ at which $V_{\rm sc}(Z_h)$ reaches its minimum, and $E_0$ is the width parameter controlling the spread around the most probable charge partition. In the present context, the essential point is that the local energetic landscape entering Eqs.~(\ref{eq:Vsc_Zh}) and (\ref{eq:Wigner_charge}) should contain a realistic finite-temperature pairing contribution if the odd--even effect is to be linked to microscopic correlations rather than introduced only at an effective level.



All available experimental data of the fission fragment charge yield are summarized in Table~\ref{tab:odd_even_statistics} obtained in the neutron (14 MeV, 0.5 MeV, and thermal) induced and spontaneous (sf) fission. The plus-sign (+) indicates that the odd-even effect is present in a given reaction, while the minus-sign (-) signals its absence. In addition, the energy differences between the ground-state and the most probable asymmetric scission point obtained in the Lublin-Strasbourg-Drop (LSD)~\cite{PomorskiDudek2003LSD} and the ISOscalar Liquid Drop Approximation (ISOLDA)~\cite{PomorskiXiao2025ISOLDA} models are displayed in the table. It shows that the odd--even effect survives only when the deformation energy and the initial excitation energy (the neutron binding and kinetic energy) at the asymmetric saddle remain below a characteristic upper limit. In the LSD model, this limit is about $23$~MeV, while in ISOLDA it increases to about $27$~MeV. The existence of such a threshold strongly suggests that odd--even staggering is controlled by the competition between intrinsic excitation and residual pairing correlations: once the available energy becomes too large, pair breaking washes out the discrete charge structure and the staggering disappears. Although the precise numerical threshold depends on the macroscopic prescription, the physical conclusion is robust. The odd--even effect, therefore, appears as an empirical signature of pairing persistence in the late stage of the fission process. One has to note also that in the cases of odd-Z parent nuclei, the odd-even effect in the fission fragment charge is weaker than in their even-even neighbors. This is well known: the odd proton weakens proton pairing correlations.

\begin{table}[t]
\caption{Statistics of the odd--even effect. The quantities labeled LSD and ISOLDA denote $\Delta E = M(\mathrm{asym.\ sciss.})-M(\mathrm{sphere})$, evaluated with the corresponding macroscopic formulae. The symbols in columns $n14$, $n05$, $n_{\rm th}$, and ${\rm sf}$ indicate the presence ({\bf +}) or absence ({\bf -}) of the odd--even effect for 14-MeV neutron-induced, 0.5-MeV neutron-induced, thermal-neutron-induced, and spontaneous fission, respectively.}
\label{tab:odd_even_statistics}
\centering
\scriptsize
\setlength{\tabcolsep}{4.5pt}
\renewcommand{\arraystretch}{1.05}
\begin{tabular}{cccccccc}
\toprule

$Z$ & $A$ & LSD & ISOLDA & $n14$ & $n05$& $n_{\rm th}$ & ${\rm sf}$ \\
&& MeV & MeV&&&&\\
\midrule
90 & 228 &8.5 & 12.6 &&& {\bf +} &\\
90 & 230 &8.0 & 12.3 &&& {\bf +} &\\
90 & 233 &7.4 & 12.0 & {\bf -} & {\bf +} &&\\
91 & 232 & 10.4 & 14.6 && + &&\\
92 & 233 & 13.1 & 17.1 &&& {\bf +} &\\
92 & 234 & 12.8 & 17.0 & {\bf -} & {\bf +} & {\bf +} &\\
92 & 235 & 12.6 & 16.8 & {\bf -} & {\bf +} &&\\
92 & 236 & 12.3 & 16.7 & {\bf -} & {\bf +} & {\bf +} &\\
92 & 237 & 12.1 & 16.6 & {\bf -} & {\bf +} &&\\
92 & 238 & 11.9 & 16.5 && {\bf +} && {\bf +} \\
92 & 239 & 11.7 & 16.5 & {\bf -} & {\bf +} &&\\
93 & 238 & 14.8 & 19.1 & {\bf -} & {\bf -} & {\bf -} &\\
93 & 239 & 14.6 & 19.0 && {\bf -} &&\\
94 & 239 & 17.5 & 21.6 && {\bf +} &&\\
94 & 240 & 17.3 & 21.5 & {\bf -} & {\bf +} & {\bf +} &\\
94 & 241 & 17.0 & 21.4 & {\bf -} & {\bf +} & {\bf +} &\\
94 & 242 & 16.8 & 21.3 && {\bf +} & {\bf +} &\\
94 & 243 & 16.6 & 21.2 & {\bf -} & {\bf +} & {\bf +} &\\
95 & 242 & 19.8 & 23.9 & {\bf -} & {\bf -} & {\bf -} &\\
95 & 243 & 19.5 & 23.8 &&& {\bf -} &\\
95 & 244 & 19.3 & 23.7 && {\bf -} &&\\
96 & 243 & 22.6 & 26.6 && {\bf -} &&\\
96 & 244 & 22.3 & 26.4 && {\bf -} & {\bf -} & {\bf +} \\
96 & 245 & 22.1 & 26.3 && {\bf -} &&\\
96 & 246 & 21.8 & 26.2 &&& {\bf -} & {\bf +} \\
96 & 247 & 21.6 & 26.0 && {\bf -} &&\\
96 & 248 & 21.4 & 25.9 &&&& {\bf -} \\
96 & 249 & 21.2 & 25.8 && {\bf -} &&\\
98 & 250 & 27.0 & 31.2 &&& {\bf -} & {\bf -} \\
98 & 252 & 26.5 & 30.9 &&& {\bf -} & {\bf -} \\
99 & 253 & 29.4 & 33.6 &&&& {\bf -} \\
99 & 255 & 28.9 & 33.4 &&& {\bf -} &\\
100 & 254 & 32.3 & 36.3 &&&& {\bf -} \\
100 & 256 & 31.8 & 36.1 &&& {\bf -} & {\bf -} \\
\bottomrule
\end{tabular}

\vspace{1ex}

\begin{minipage}{0.98\columnwidth}
\end{minipage}
\end{table}

According to the methodology presented above, solving the coupled Langevin equations described in Sec.~\ref{sec:method} yields distributions of masses and shapes of nearly touching prefragments at the scission configuration. To convert these scission configurations into fragment charge yields, an additional charge-equilibration mechanism must be introduced. The corresponding prescription, developed in Ref.~\cite{Pomorski2023FoS}, is outlined in Sec.~\ref{sec:cheq}.

 It was shown in Ref.~\cite{PomorskiAcceptedBaXe} that this method reproduces rather well a global structure of the fission fragment charge yields in different regions of fissioning nuclei. A crucial element to reproduce the value of the odd-even effect in the fission fragment charge yields is the pairing gap value in each emerging fragment. The pairing gap diminishes with temperature. So, the pairing gap, which reproduces the experimental odd-even effect well, may serve as a kind of thermometer, i.e., it provides important information about the magnitude of energy dissipation in a given fission process.
\begin{figure}[t]
\centering
\includegraphics[width=1.0\columnwidth]{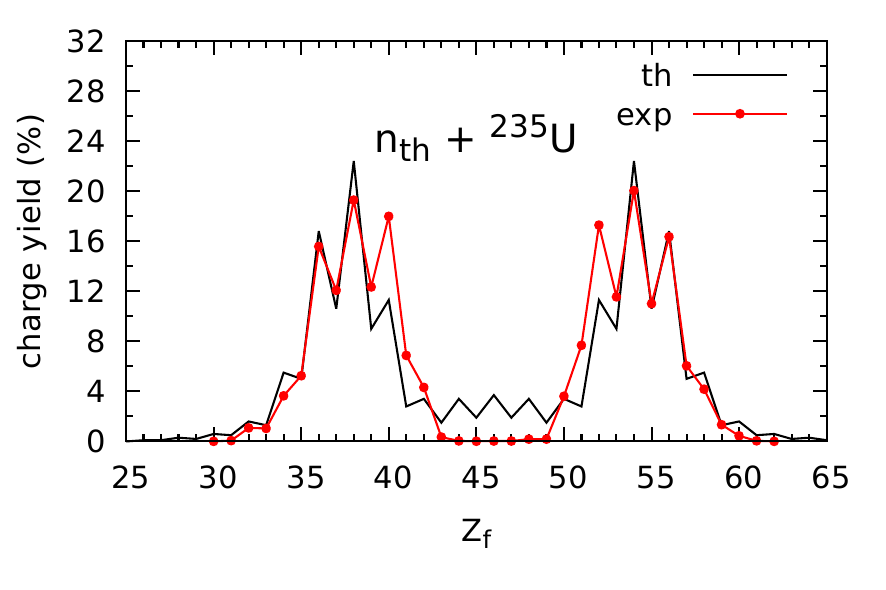}
\caption{Calculated fragment charge yield for thermal-neutron-induced fission of $^{235}$U illustrating the odd--even staggering generated by the charge-polarization prescription at scission. The visible alternation between neighboring even and odd fragment charges suggests that a residual pairing-related structure survives into the final charge partition.}
\label{fig:U236th_chy}
\end{figure}

Figure~\ref{fig:U236th_chy} taken from Ref.~\cite{Pomorski2023FoS} shows that the calculated charge-yield distribution reproduces the gross structure of the experimental data rather well. In particular, the positions of the two dominant asymmetric maxima are correctly determined, indicating that the average charge partition at scission is described realistically. The odd--even staggering is also clearly visible in both theory and experiment, which shows that the charge-polarization prescription is able to generate the essential alternating pattern between neighboring even and odd fragment charges. The predicted odd-even staggering is slightly larger than the experimental one because we have used a standard pairing gap for the fragment in the cold systems.

At the same time, some deficiencies remain at a more differential level. Although the staggering is reproduced, the theory does not recover the detailed peak structure with sufficient accuracy on the more symmetric side of the main asymmetric maxima, nor in the distribution's asymmetric tails. In addition, the calculation produces a weak symmetric component that is essentially absent in the data. This suggests that the theoretical population of the symmetric valley is still somewhat too large, most likely due to the interplay between fluctuations and dissipation in the immediate vicinity of the scission line.

A detailed analysis of the odd-even effect in yields across different fissioning nuclei and excitation energies may provide valuable insights into the pairing strength at large deformations. Already now, it is clear that the constant, deformation-independent pairing strength is unable to reproduce odd-even effects.


\section{Discussion and conclusions}
\label{sec:discussion_conclusions}

The present shows that the role of pairing changes significantly with deformation. Near the ground state, the constant-$G$ and surface-dependent prescriptions give results that are relatively similar, but their difference increases with elongation and reaches a maximum in the scission region. For the pairing free-energy correction, this difference is substantial: in the constant-$G$ prescription, the pairing contribution becomes very small at large deformation, whereas in the surface-dependent prescription it remains non-negligible and, in the symmetric scission valley, becomes particularly strong, especially for neutrons. This means that the pairing properties in the scission region cannot be inferred solely from compact configurations.

The results obtained in this work show that the odd--even effect in fragment charge yields can be related to the finite-temperature survival of pairing correlations in the strongly deformed pre-scission system. The analysis of the pairing gap, the pairing free-energy correction, and the shell free-energy correction indicates that the microscopic structure is not uniformly suppressed as temperature increases. Instead, both pairing and shell terms follow regular attenuation patterns that remain meaningful from compact shapes up to scission-like configurations. The pairing gap and the pairing contribution to the free energy can be represented by compact scaling laws, while the shell free-energy correction follows a damping pattern of Bohr--Mottelson type governed mainly by the local shell spacing.

The shell contribution shows a similarly systematic evolution. At low temperature, the shell free-energy correction around scission exhibits pronounced minima associated with both asymmetric and nearly symmetric configurations. With increasing temperature, the fine oscillatory structure disappears first, the main minima become shallower and broader, and only at higher temperatures does the map become nearly flat. This implies that shell effects are reduced progressively rather than abruptly. At intermediate excitation energies, the detailed shell structure is already weakened, but the broader bias between competing valleys may still remain.

The charge-yield calculations are consistent with this picture. The positions of the dominant asymmetric maxima are reproduced correctly, and the odd--even staggering is clearly present in the calculated distributions. This indicates that the scission-point charge-polarization mechanism captures the main features of the integer charge partition. At the same time, some discrepancies remain: the local peak amplitudes are not reproduced accurately enough on the more symmetric side of the main asymmetric maxima and in the asymmetric tails, and the calculation produces a weak symmetric component that is not seen in the data. These differences suggest that the treatment of fluctuations and dissipation in the immediate vicinity of the scission line still requires refinement.

Within the macroscopic--microscopic framework, the deformation-dependent energy is written as a sum of a smooth macroscopic term and microscopic shell and pairing corrections. In the present discussion, the macroscopic contribution is considered in two variants, based on the Lublin--Strasbourg Drop (LSD) model~\cite{PomorskiDudek2003LSD} and the ISOscalar Liquid Drop Approximation (ISOLDA)~\cite{PomorskiXiao2025ISOLDA}, in order to test the robustness of the conclusions with respect to the choice of the underlying smooth energy.

The odd--even statistics lead to an additional conclusion. The effect survives only when the total available energy, i.e., the sum of the deformation and the initial excitation energies, at the asymmetric saddle remains below a characteristic upper limit, about 23 MeV in the LSD description and about 27 MeV in ISOLDA. Although the precise threshold depends on the macroscopic prescription, the existence of such a limit appears stable. This indicates that odd--even staggering is controlled by the competition between intrinsic excitation and residual pairing correlations: once the energy available for pair breaking becomes too large, the discrete charge structure is no longer maintained.

These results also have a methodological consequence. In earlier isotope-resolved Langevin calculations, the temperature dependence of the microscopic contribution was introduced through an effective damping factor applied to the total microscopic energy. The present analysis shows that this treatment is not sufficient. Pairing and shell corrections differ not only in magnitude, but also in deformation dependence and in thermal attenuation pattern. They should therefore be treated separately in finite-temperature transport calculations. In practical terms, this means that future Langevin simulations of charge yields, fragment observables, and neutron-emission systematics should include explicit deformation- and temperature-dependent pairing and shell terms in the free energy, rather than a single common damping prescription.

From a methodological perspective, the present study also introduces a set of simple, nearly universal analytical formulas for the thermal attenuation of pairing and shell effects. These parametrizations may be particularly useful for practical finite-temperature fission calculations. In particular, the damping of the shell free-energy correction is shown to follow a Bohr--Mottelson-type behavior governed mainly by the local shell spacing.

Overall, the present work provides a finite-temperature microscopic basis for interpreting odd-even staggering in fission-fragment charge yields. It shows that the observed charge structure can be understood as a manifestation of the persistence and thermal damping of pairing and shell effects in the scission region, and it identifies the proper treatment of these contributions as an essential ingredient of realistic dynamical calculations of fission observables. In particular, the systematic behavior of thermal-neutron-induced charge yields across the actinide region supports this interpretation: the odd--even staggering is strongest in the lighter systems, where pairing correlations still leave a pronounced imprint on the final charge partition, and gradually weakens towards heavier fissioning nuclei, where the charge distributions become smoother and the effect is strongly reduced or nearly absent. This trend indicates that the magnitude of odd--even staggering is not universal, but depends sensitively on the fissioning nucleus and, indirectly, on the intrinsic excitation carried by the system at large deformation. The weaker staggering observed for odd-$Z$ parent nuclei provides further support for this picture, being consistent with the expected reduction of proton pairing correlations in the presence of an unpaired proton. A systematic reproduction of these experimentally observed effects within the framework presented here will be the subject of future investigations.
\begin{acknowledgments}
The authors are sincerely grateful to F.~A.~Ivanyuk for valuable discussions and many helpful suggestions. They also express their sincere thanks to B. Nerlo - Pomorska for a very careful and detailed analysis of the manuscript and for numerous insightful remarks.
This research was funded in part by the National Science Centre, Poland, under Project No. 2023/49/B/ST2/01294. It was also supported by the Natural Science Foundation of China under Grant Nos. 11961131010 and 12335008. In addition, M. Kowal and T. Cap were partially supported by the Polish-French cooperation COPIGAL. 
\end{acknowledgments}


\bibliographystyle{apsrev4-2}

\bibliography{pairing}

\end{document}